\documentclass[showpacs,preprintnumbers,twocolumn,superscriptaddress,aps,nofootinbib]{revtex4}
\usepackage{amssymb}
\usepackage[centertags]{amsmath}
\usepackage{txfonts}
\usepackage{epsfig}
\usepackage{bm}
\usepackage{color}
\usepackage{graphicx,graphics}
\usepackage{multirow}
\usepackage{float}
\usepackage[pdfstartview=FitH]{hyperref}
\hypersetup{colorlinks=true, citecolor=blue,linkcolor=blue,filecolor=black,urlcolor=blue}
\allowdisplaybreaks[2]

\begin{document}

\title{Hadron Yield Correlation in Quark Combination Models in High-Energy $AA$ Collisions}

\author{Rui-qin Wang}
\affiliation{School of Physics, Shandong University, Jinan, Shandong 250100, China}

\author{Feng-lan Shao}
\affiliation{Department of Physics, Qufu Normal University, Shandong 273165, China}

\author{Jun Song}
\affiliation{School of Physics, Shandong University, Jinan, Shandong 250100, China}

\author{Qu-bing Xie}
\affiliation{School of Physics, Shandong University, Jinan, Shandong 250100, China}

\author{Zuo-tang Liang}
\affiliation{School of Physics, Shandong University, Jinan, Shandong 250100, China}

\begin{abstract}
We study the hadron yield correlation in the combination models in high-energy heavy-ion collisions.
We derive the relationship between the average yields of different hadrons produced in the
combination of a system consisting of equal numbers of quarks and antiquarks.
We present the results for the directly produced hadrons as well as those for the
final hadrons including the strong and electromagnetic decay contributions.
We also study the net quark influence by considering the case when the number of quarks is larger than 
that of antiquarks. We make comparison with the data wherever possible.
\end{abstract}

\maketitle

\section{introduction}

Hadron yield correlations, measured by the ratios of the average yields of different hadrons produced in high-energy reactions,
are one kind of characteristic properties of hadronization mechanisms.
It is usually expected that these correlations are more or less independent of the particular model especially
when the combination mechanism is concerned.
Such properties were therefore considered as a good probe for the hadronization mechanism in different high-energy reactions already
in the 1970s to 1990s \cite{Anisovich1973NPB,Bjorken1974PRD,sdpi1981PRD,1982PRD,Liang1991PRD,thermal1995PLB,Sa1995PRC,Bialas1998PLB}.
The study of these correlations has attracted much attention in heavy-ion collisions 
recently \cite{Zimanyi2000PLB,Greco2003PRL,Fries2003PRL,RCHwa2003PRC,
FLShao2005PRC,TYao2008PRC,CEShao2009PRC}
because they are considered as a probe to test whether the (re-)combination mechanism is at work.
This is interesting because whether the combination mechanism works might be considered as one of the signatures for the formation
of the bulk color-deconfined quark matter system before the hadronization takes place.
Experimental results are available from the Relativistic Heavy Ion Collider (RHIC) \cite{starESC,BES2012nuclex}, 
from relatively low-energy collisions such as those obtained by NA49, NA61, and CBM Collaborations 
at the Super Proton Synchrotron (SPS) \cite{BES2012nuclex,20A30A2008PRC,NA49bes2011nuclex,CBM2009nuclex}, 
and more recently from the very high energy reactions at the Large Hadron Collider (LHC) \cite{pikp2011nuclex,sBpi2012nuclex}.
These results seem to suggest a dramatic change for collisions from lower to higher energies,
which is considered as one of the hints for  phase transition \cite{20A30A2008PRC}.

In order to make a judgment whether the combination mechanism is at work by comparing
the theoretical results with the corresponding experimental data, it is important to see whether, if so,
to what extent, the theoretical results depend on the particular model used in obtaining these results.
There are many studies that have been made in the
literature \cite{Zimanyi2000PLB,Greco2003PRL,Fries2003PRL,RCHwa2003PRC,FLShao2005PRC,FLShao2007PRC}.
These studies are usually based on some particular (re-)combination or coalescence models and/or some particular assumptions.
It is not clear whether the results obtained depend on the particular assumption(s) made in these particular models.
For this purpose, in this paper, we will make a systematic study of the average yields of different identified hadrons and
their relationships obtained in the combination mechanism.
We will start the study by considering the case for the combination of a system of
quarks and antiquarks from the basic ideas of the combination mechanism.
We will make the study as independent of the particular models as possible but present the assumptions and/or inputs explicitly 
whenever necessary.

The rest of the paper is organized as follows.
In Sec.~II, we derive the formulas for calculating the average yields of hadrons and their relationships
in the combination of a system of quarks and antiquarks.
We consider a system where the number of quarks and that of antiquarks are equal and
discuss the net quark influence as well.
We compare the results with the available data in Sec.~III.
Here, the experimental results from the LHC \cite{pikp2011nuclex,sBpi2012nuclex} are taken as an example 
to test the predictions for the case where the net quark influence is considered as negligible, 
while those from the RHIC and the SPS 
\cite{20A30A2008PRC,plxoPRCSPS,plxoPRC62200,pPRC200,pPRC130,lxoPLB130,LPRL130,PLPRC2006,hypRL200,LPRL1588040,
MULSPRL130,OSPRL15840,PPLB62,kPRC4080158} are used to test the net quark influence.
A short summary is presented in Sec.~V.

\section{Hadron yield ratios in the combination models}

In this section, we begin with the general formalism of hadron yields in
the combination models based on the basic ideas.
For this purpose, we start with a quark-antiquark
system as general as possible.
Then we simplify the results by using some explicit assumptions, simplifications, and/or approximations.
We present the results for the ratios of the yields of hadrons directly produced as well as those including
the contributions from the resonance decays.

\subsection{The general formalism}

We start with the most general case and consider a system of $N_q$ quarks and $N_{\bar q}$ antiquarks. 
We denote the number of quarks of flavor $q_i$ by $N_{q_i}$ so that $\sum_i N_{q_i}=N_q$ and 
similarly $\sum_i N_{\bar q_i}=N_{\bar q}$. 
These quarks and antiquarks combine with each other to form the color singlet hadrons.
The number density of the directly produced hadrons is given by
{\setlength\arraycolsep{0.2pt}
\begin{eqnarray}
 f_{M_j}(p_{M_{j}})= & \sum\limits_{q_{1}\bar{q}_{2}}  & \int dp_{1} dp_{2} 
 f_{q_{1}\bar{q}_{2}}(p_{1},p_{2};N_{q_i},N_{\bar{q}_i})\times        \nonumber\\
 &&  \mathcal {R}_{M_j,q_{1}\bar{q}_{2}}(p_{M_j},p_{1},p_{2};N_{q_i},N_{\bar{q}_i}), \label{eq:Mgenal} \\
 f_{B_j}(p_{B_{j}})=&\sum\limits_{q_{1}q_{2}q_{3}} & \int dp_1 dp_2 dp_3    \times \nonumber\\
 && f_{q_{1}q_{2}q_{3}}(p_1,p_2,p_3;N_{q_i},N_{\bar{q}_i})     \times \nonumber \\
 && \mathcal {R}_{B_j,q_{1}q_{2}q_{3}}(p_{B_j},p_1,p_2,p_3;N_{q_i},N_{\bar{q}_i}),
\label{eq:Bgenal}
\end{eqnarray} }where $f_{M_j}$ and $f_{B_j}$ 
are the momentum distributions for the produced meson $M_j$ and baryon $B_j$, respectively;
$f_{q_{1}\bar{q}_{2}}(p_1,p_2;N_{q_i},N_{\bar{q}_i})$ and
$f_{q_{1}q_{2}q_{3}}(p_1,p_2,p_3;N_{q_i},N_{\bar{q}_i})$ are the two- and three-particle joint
momentum distributions for $(q_{1}\bar{q}_{2})$ and $(q_{1}q_{2}q_{3})$, respectively.
The kernel functions $\mathcal {R}_{M_j,q_{1}\bar{q}_{2}}(p_{M_j},p_1,p_2;N_{q_i},N_{\bar{q}_i})$ and
$\mathcal {R}_{B_j,q_{1}q_{2}q_{3}}(p_{B_j},p_1,p_2,p_3;N_{q_i},N_{\bar{q}_i})$
stand for the probability density for $q_1$ and $\bar{q}_2$ with momenta
$p_1$ and $p_2$ to combine into a meson $M_j$ of momentum $p_{M_j}$
and that for $q_1$, $q_2$ and $q_3$ with momenta $p_1$, $p_2$, and $p_3$ to
coalescence into a baryon $B_j$ of momentum $p_{B_j}$. 
Here, in the arguments, we use $N_{q_i}$ and $N_{\bar{q}_i}$ to represent the dependence of these functions on
the numbers of the quarks and antiquarks of different flavors, and also on the total collision energy $\sqrt{s}$ of $AA$ reactions.
We note in particular that not only the joint distributions $f_{q_{1}\bar{q}_{2}}$ and
$f_{q_{1}q_{2}q_{3}}$ but also the probability densities $\mathcal {R}_{M_j,q_{1}\bar{q}_{2}}$
and $\mathcal {R}_{B_j,q_{1}q_{2}q_{3}}$ are in general dependent on $N_{q_i}$ and $N_{\bar{q}_i}$. 
This is because, for finite $N_{q_i}$ and $N_{\bar{q}_i}$, the probability for a given quark $q_1$ to combine with 
a specified antiquark $\bar q_2$ to form a specified meson $M_j$ or two other specified quarks $q_2q_3$ to 
form a baryon $B_j$ is in general dependent on the number $N_{q_i}$ of existing quarks of different flavors 
and the number $N_{\bar{q}_i}$ of antiquarks.

We note in particular the relationship between the description presented here and those given in the literature in 
different models based on the combination mechanism such
as the coalescence model  \cite{Zimanyi2000PLB,Greco2003PRL}, the recombination model \cite{RCHwa2003PRC,Fries2003PRL}, 
and the quark combination model \cite{QBXie1988PRD,FLShao2005PRC,CEShao2009PRC} developed by different groups.
Eqs.~(\ref{eq:Mgenal}) and (\ref{eq:Bgenal}) are intended to be the general formulas based on the basic
ideas of the combination mechanism. The different models
are examples of the general case that we considered in these
equations. In these models, different method(s) and/or assumption(s) are usually introduced to construct the precise
form of the kernel functions
$\mathcal {R}_{M_j,q_{1}\bar{q}_{2}}(p_{M_j},p_1,p_2;N_{q_i},N_{\bar{q}_i})$ and
$\mathcal {R}_{B_j,q_{1}q_{2}q_{3}}(p_{B_j},p_1,p_2,p_3;N_{q_i},N_{\bar{q}_i})$ in order to provide a good
description of different properties of the hadrons, such as momentum distributions and so on.
For example, in the recombination model developed by Hwa and collaborators \cite{RCHwa2003PRC}, 
these kernel functions are just the recombination functions. 

The joint distributions $f_{q_{1}\bar{q}_{2}}$ and $f_{q_{1}q_{2}q_{3}}$  are the number densities that satisfy
\begin{eqnarray}
&\int f_{q_{1}\bar{q}_{2}}(p_1,p_2;N_{q_i},N_{\bar{q}_i}) dp_1 dp_2=N_{q_{1}\bar{q}_2},  \\
&\int f_{q_{1}q_{2}q_{3}}(p_1,p_2,p_3;N_{q_i},N_{\bar{q}_i}) dp_1 dp_2 dp_3=N_{q_1q_2q_3}, \ \ \ 
\end{eqnarray} 
respectively, where $N_{q_{1}\bar{q}_2}=N_{q_1}N_{\bar{q}_2}$, and
\begin{equation}
N_{q_1q_2q_3} = \left\{ \begin{array}{ll}
N_{q_1}N_{q_2}N_{q_3} & \textrm{for    } q_1 \neq q_2 \neq q_3\\
N_{q_1}(N_{q_1}-1)N_{q_3} & \textrm{for    } q_1=q_2 \neq q_3\\
N_{q_1}(N_{q_1}-1)(N_{q_1}-2) & \textrm{for    } q_1=q_2=q_3
\end{array} \right.
\end{equation}
are the numbers of all the possible $(q \bar q)$'s and $(qqq)$'s in the bulk quark-antiquark system that
we consider.
For the convenience of comparison, we rewrite them as
{\setlength\arraycolsep{1pt}
\begin{eqnarray}
 &&f_{q_{1}\bar{q}_{2}}(p_1,p_2;N_{q_i},N_{\bar{q}_i})=N_{q_{1}\bar{q}_2} f^{(n)}_{q_{1}\bar{q}_{2}}(p_1,p_2;N_{q_i},N_{\bar{q}_i}), \\
 &&f_{q_{1}q_{2}q_{3}}(p_1,p_2,p_3;N_{q_i},N_{\bar{q}_i})=
                     \nonumber     \\
 &&~~~~~~~~~~~~~~~~~~~~~~~~~~~~N_{q_1q_2q_3} f^{(n)}_{q_{1}q_{2}q_{3}}(p_1,p_2,p_3;N_{q_i},N_{\bar{q}_i}),~~~~~~
\end{eqnarray}}so that the distributions are normalized to unity where we denote by using the superscript ($n$), i.e., 
{\setlength\arraycolsep{1pt}
\begin{eqnarray}
 &\int& f^{(n)}_{q_{1}\bar{q}_{2}}(p_1,p_2;N_{q_i},N_{\bar{q}_i}) dp_1 dp_2=1,   \\
 &\int& f^{(n)}_{q_{1}q_{2}q_{3}}(p_1,p_2,p_3;N_{q_i},N_{\bar{q}_i}) dp_1 dp_2 dp_3=1. ~~
\end{eqnarray}}In terms of these normalized joint distributions, we have
{\setlength\arraycolsep{1pt}
\begin{eqnarray}
 && f_{M_j}(p_{M_j})=\sum \limits_{q_{1}\bar{q}_{2}} N_{q_1 \bar{q}_2} \int dp_1 dp_2
 f^{(n)}_{q_{1}\bar{q}_{2}}(p_1,p_2;N_{q_i},N_{\bar{q}_i}) \nonumber \\
 && ~~~~~~~~~~~~~~~~~~~~~~~~~~~~~~~ \times  \mathcal {R}_{M_j}(p_{M_j},p_1,p_2;N_{q_i},N_{\bar{q}_i}), \\
 && f_{B_j}(p_{B_j})=\sum \limits_{q_{1}q_{2}q_{3}} N_{q_1q_2q_3} \int dp_1 dp_2 dp_3 \times \nonumber \\
 &&  ~~~~~~~~~~~~~~~~~~~~~~~~~~~~~~~  f^{(n)}_{q_{1}q_{2}q_{3}}(p_1,p_2,p_3;N_{q_i},N_{\bar{q}_i})  \times \nonumber \\
 &&  ~~~~~~~~~~~~~~~~~~~~~~~~~~~~~~~  \mathcal {R}_{B_j}(p_{B_j},p_1,p_2,p_3;N_{q_i},N_{\bar{q}_i}).~~~~~~
\end{eqnarray} }Integrating over $p_{M_j}$ or $p_{B_j}$ from the momentum distributions, we obtain the average numbers of the directly produced mesons $M_j$ and baryons $B_j$ as
{\setlength\arraycolsep{0.5pt}
\begin{eqnarray}
 &&\overline{N}_{M_j}(N_{q_i},N_{\bar{q}_i}) =\sum \limits_{q_{1}\bar{q}_{2}} N_{q_1 \bar{q}_2}
 \int dp_{M_j} dp_1 dp_2    \times
 \nonumber  \\
&& ~~~~~~~  f^{(n)}_{q_{1}\bar{q}_{2}}(p_1,p_2;N_{q_i},N_{\bar{q}_i})
  \mathcal {R}_{M_j}(p_{M_j},p_1,p_2;N_{q_i},N_{\bar{q}_i}),~~~~ \\
 &&\overline{N}_{B_j}(N_{q_i},N_{\bar{q}_i})
     =\sum \limits_{q_{1}q_{2}q_{3}} N_{q_1q_2q_3} \int dp_{B_j} dp_1 dp_2 dp_3 \times \nonumber \\
&& f^{(n)}_{q_{1}q_{2}q_{3}}(p_1,p_2,p_3;N_{q_i},N_{\bar{q}_i}) \mathcal {R}_{B_j}(p_{B_j},p_1,p_2,p_3;N_{q_i},N_{\bar{q}_i}).
     \nonumber  \\
\end{eqnarray} }For a reaction at a given energy, the average numbers of quarks, $\langle N_{q_i}\rangle$, and those for the
antiquarks, $\langle N_{\bar q_i}\rangle$, of different flavors are fixed. The numbers of quarks and antiquarks follow a certain distribution which we denote by
$P(N_{q_i},N_{\bar q_i},\langle N_{q_i}\rangle,\langle N_{\bar q_i}\rangle)$. The average yields of mesons and baryons are given by
{\setlength\arraycolsep{0.5pt}
\begin{eqnarray}
&&\langle N_{M_j}\rangle(\sqrt{s}~)=\sum_{N_{q_i}N_{\bar q_i}} P(N_{q_i},N_{\bar q_i},\langle N_{q_i}\rangle,\langle N_{\bar q_i}\rangle)\overline{N}_{M_j}(N_{q_i},N_{\bar{q}_i}),
        \nonumber  \\
&&\langle N_{B_j}\rangle(\sqrt{s}~)=\sum_{N_{q_i}N_{\bar q_i}} P(N_{q_i},N_{\bar q_i},
\langle N_{q_i}\rangle,\langle N_{\bar q_i}\rangle)
\overline{N}_{B_j}(N_{q_i},N_{\bar{q}_i}). \nonumber
\end{eqnarray} }

These equations are the general formalism for calculating the average yield of a certain sort of hadrons in high-energy reactions
based on the basic ideas of the combination mechanism.
More specific results can be obtained for special cases when special assumptions are made about the distributions and/or the kernel functions.
We present such cases step by step in the following.

\subsection{Factorization of flavor and momentum dependences}

The flavor dependence of the kernel functions $\mathcal {R}_{M_j,q_{1}\bar{q}_{2}}$ and $\mathcal {R}_{B_j,q_{1}q_{2}q_{3}}$
is responsible for flavor conservation in the combination process and the differences between the combination
probabilities for different flavors of quarks, antiquarks, and hadrons.
In general, the momentum and the flavor dependencies of these kernel functions are coupled to each other.
In that case, the results for the ratios of the average yields of different hadrons can be dependent on the way of coupling. 
In this paper, we do not consider such coupling effects.  
In contrast, in the following, we consider only the simplest case 
where the momentum and flavor dependencies of the kernel functions are decoupled from each other. 
In other words, we consider the case where they are factorized, i.e.,
{\setlength\arraycolsep{0.5pt}
\begin{eqnarray}
 &&\mathcal {R}_{M_j,q_{1}\bar{q}_{2}}(p_{M_j},p_1,p_2;N_{q_i},N_{\bar{q}_i})=   \nonumber \\
 && ~~~~~~
 \mathcal {R}^{(f)}_{M_j,q_{1}\bar{q}_{2}}(N_{q_i},N_{\bar{q}_i})
 \mathcal {R}^{(p)}_{M}(p_{M},p_1,p_2;N_q,N_{\bar q}), \\
 &&\mathcal {R}_{B_j,q_{1}q_{2}q_{3}}(p_{B_j},p_1,p_2,p_3;N_{q_i},N_{\bar{q}_i})=    \nonumber \\
 && ~~~~~~
 \mathcal {R}^{(f)}_{B_j,q_{1}q_{2}q_{3}}(N_{q_i},N_{\bar{q}_i})
 \mathcal {R}^{(p)}_{B}(p_{B},p_1,p_2,p_3;N_q,N_{\bar q}), ~~~~~~
\end{eqnarray}}where the flavor-independent parts $\mathcal {R}^{(p)}_{M}(p_{M},p_1,p_2;N_q,N_{\bar q})$ and 
$\mathcal {R}^{(p)}_{B}(p_{B},p_1,p_2,p_3;N_q,N_{\bar q})$ denote the probability for a $(q \bar q)$ with 
momenta $p_1$ and $p_2$ in a system consisting of $N_q$ quarks and $N_{\bar q}$ antiquarks to combine with
each other to form a meson $M$ with momentum $p_M$ and that for a $(qqq)$ with 
momenta $p_1$, $p_2$, and $p_3$ in the system to combine with
each other to form a baryon $B$ with momentum $p_B$, respectively.
The flavor-dependent parts $\mathcal {R}^{(f)}_{M_j,q_1\bar q_2}(N_{q_i},N_{\bar{q}_i})$ and
$\mathcal {R}^{(f)}_{B_j,q_{1}q_{2}q_{3}}(N_{q_i},N_{\bar{q}_i})$ represent the probability
for the $q_1$ and $\bar q_2$ to combine into the specified meson $M_j$ in the case that they are known to
combine into a meson and that for the $q_1$, $q_2$, and $q_3$ to combine into the specified baryon $B_j$
in the case that they are known to combine into a baryon, respectively.
They are taken as satisfying the normalization condition
\begin{eqnarray}
&& \sum_j   \mathcal {R}^{(f)}_{M_j,q_1\bar q_2} =1, \\
&& \sum_j   \mathcal {R}^{(f)}_{B_j,q_{1}q_{2}q_{3}} =1.
\end{eqnarray}
We further assume that the normalized joint momentum distributions of the quarks and/or antiquarks are flavor independent, i.e.,
{\setlength\arraycolsep{1pt}
\begin{eqnarray}
 &&f^{(n)}_{q_{1} \bar{q}_{2}}(p_1,p_2;N_{q_i},N_{\bar{q}_i})=
 f^{(n)}_{q \bar{q}}(p_1,p_2;N_{q},N_{\bar{q}}), \\
&&f^{(n)}_{q_1q_2q_3}(p_1,p_2,p_3;N_{q_i},N_{\bar{q}_i})=
 f^{(n)}_{qqq}(p_1,p_2,p_3;N_{q},N_{\bar{q}}). \ \ \ \ \ \ \ \
\end{eqnarray}}
Under these two approximations, we have
{\setlength\arraycolsep{0.5pt}
\begin{eqnarray}
 &&\overline{N}_{M_j}(N_{q_i},N_{\bar{q}_i})=
 \sum \limits_{q_{1}\bar{q}_{2}} N_{q_1 \bar{q}_2} \mathcal {R}^{(f)}_{M_j,q_{1}\bar{q}_{2}}(N_{q_i},N_{\bar{q}_i}) \times
                               \nonumber \\
 &&~~~~~~~~~~~~~~~~~~~~~\int dp_{M} dp_1 dp_2 f^{(n)}_{q \bar{q}}(p_1,p_2;N_{q},N_{\bar{q}}) \times~~~~~~
                                \nonumber   \\
 &&~~~~~~~~~~~~~~~~~~~~~~~~~~~\mathcal {R}^{(p)}_{M}(p_{M},p_1,p_2;N_q,N_{\bar q}),~~~  \\[0.3cm]
 && \overline{N}_{B_j}(N_{q_i},N_{\bar{q}_i})=\sum \limits_{q_{1}q_{2}q_{3}} N_{q_1q_2q_3} \mathcal {R}^{(f)}_{B_j,q_{1}q_{2}q_{3}}(N_{q_i},N_{\bar{q}_i})
                             \times \nonumber  \\
 && ~~~~~~~~~~~~~~~~\int dp_{B} dp_1 dp_2 dp_3
 f^{(n)}_{qqq}(p_1,p_2,p_3;N_{q},N_{\bar{q}}) \times\nonumber  \\
 && ~~~~~~~~~~~~~~~~~~~~~~ \mathcal {R}^{(p)}_{B}(p_{B},p_1,p_2,p_3;N_q,N_{\bar q}).
\end{eqnarray} }We denote
{\setlength\arraycolsep{0.5pt}
\begin{eqnarray}
 &&\gamma_M(N_{q},N_{\bar{q}},\sqrt{s})=\int dp_{M}dp_1 dp_2 \times          \nonumber \\
 && ~~~~~f^{(n)}_{q \bar{q}}(p_1,p_2;N_{q},N_{\bar{q}})  \mathcal {R}^{(p)}_{M}(p_{M},p_1,p_2;N_q,N_{\bar q}), \label{eq:gammaM}\\
 &&\gamma_B(N_{q},N_{\bar{q}},\sqrt{s})=\int dp_{B} dp_1 dp_2 dp_3 \times      \nonumber \\
 && ~~~~~f^{(n)}_{qqq}(p_1,p_2,p_3;N_{q},N_{\bar{q}}) \mathcal {R}^{(p)}_{B}(p_{B},p_1,p_2,p_3;N_q,N_{\bar q}), ~~~~~~\label{eq:gammaB}
\end{eqnarray} }and obtain
{\setlength\arraycolsep{0.5pt}
\begin{eqnarray}
 &&\overline{N}_{M_j}(N_{q_i},N_{\bar{q}_i})=\sum \limits_{q_{1}\bar{q}_{2}} N_{q_1 \bar{q}_2}
  \mathcal {R}^{(f)}_{M_j,q_{1}\bar{q}_{2}} \gamma_{M},  \\
 &&\overline{N}_{B_j}(N_{q_i},N_{\bar{q}_i})=\sum \limits_{q_{1}q_{2}q_{3}} N_{q_1q_2q_3} 
 \mathcal {R}^{(f)}_{B_j,q_{1}q_{2}q_{3}}\gamma_{B} .
\end{eqnarray} }Summing over different species of mesons and those of the baryons, respectively,
we obtain the average total numbers of mesons and baryons produced in the combination of the system
of $N_q$ quarks and $N_{\bar q}$ antiquarks as
{\setlength\arraycolsep{1pt}
\begin{eqnarray}
 &&\overline{N}_{M}(N_{q},N_{\bar{q}},\sqrt{s}~)= N_{q\bar q} \gamma_{M} (N_{q},N_{\bar{q}},\sqrt{s}~), \label{NM} \\
 &&\overline{N}_{B}(N_{q},N_{\bar{q}},\sqrt{s}~)=N_{qqq} \gamma_{B} (N_{q},N_{\bar{q}},\sqrt{s}~),
\end{eqnarray} }where $N_{q\bar{q}}=N_qN_{\bar q}$ and $N_{qqq}=N_q(N_q-1)(N_q-2)$ are
the total number of $q\bar q$ pairs and that of $qqq$ systems, respectively.
The factors $\gamma_{M} (N_{q},N_{\bar{q}},\sqrt{s})$ and $\gamma_{B} (N_{q},N_{\bar{q}},\sqrt{s})$ represent the
probability for a particular $q\bar q$ from the system consisting of
$N_q$ quarks and $N_{\bar q}$ antiquarks to combine with each other
to form a meson and that for a $qqq$ to form a baryon, respectively. We emphasize in particular that Eq.~(\ref{NM}) does not mean
that the average yield of mesons is proportional to the product
of the number of quarks and that of antiquarks since the
factor $\gamma_M$ can depend strongly on $N_q$ and $N_{\bar q}$. This is because,
for a given q, the larger $N_q$ and/or $N_{\bar q}$, the more possibilities
for the q to combine with the others to form a hadron, and
thus the smaller the probability for it to combine with the
given $\bar q$ to form the meson. In fact, it can in general be expected that $\gamma_M$ should be more or less inversely proportional
to $N_q$ and/or $N_{\bar q}$ and the final result for $\langle N_M\rangle$ should be roughly
proportional to $N_q+N_{\bar q}$. A similar conclusion holds for $\langle N_B\rangle$.

In terms of the total average numbers of mesons and baryons, the average number of
a specified meson $M_j$ and that of a specified baryon $B_j$ are given by
{\setlength\arraycolsep{1pt}
\begin{eqnarray}
 \overline{N}_{M_j}(N_{q_i},N_{\bar{q}_i})=&&
 \sum_{q_{1}\bar{q}_{2}} \frac{N_{q_1 \bar{q}_2}}{N_{q \bar{q}}}
 \mathcal {R}^{(f)}_{M_j,q_{1}\bar{q}_{2}}(N_{q_i},N_{\bar{q}_i})  \nonumber    \\
 && \times \overline{N}_{M}(N_{q},N_{\bar{q}},\sqrt{s}~),  \\
 \overline{N}_{B_j}(N_{q_i},N_{\bar{q}_i})=&&
 \sum_{q_{1}q_{2}q_{3}} \frac{N_{q_1q_2q_3}}{N_{qqq}}\mathcal {R}^{(f)}_{B_j,q_{1}q_{2}q_{3}}(N_{q_i},N_{\bar{q}_i})    \nonumber    \\
 &&\times \overline{N}_{B}(N_{q},N_{\bar{q}},\sqrt{s}~).
\end{eqnarray} }The parts describing the flavor dependence of the kernel functions,
$\mathcal {R}^{(f)}_{M_j,q_{1}\bar{q}_{2}}(N_{q_i},N_{\bar{q}_i})$ and 
$\mathcal {R}^{(f)}_{B_j,q_{1}q_{2}q_{3}}(N_{q_i},N_{\bar{q}_i})$,
have to guarantee flavor conservation in the combination process.
Hence, they contain the Kronecker $\delta$'s and constant factors $C_{M_j}$ and $C_{B_j}$.
For example, for $\pi^{+}$ and $p$, they are given by
{\setlength\arraycolsep{0.2pt}
\begin{eqnarray}
&& \mathcal {R}^{(f)}_{\pi^{+},q_{1}\bar{q}_{2}}=C_{\pi^+}\delta_{q_{1},u}\delta_{\bar{q}_2,\bar{d}},
           \nonumber\\
&& \mathcal {R}^{(f)}_{p,q_{1}q_{2}q_{3}}=C_{p} \bigl( \delta_{q_{1},u}\delta_{q_{2},u}\delta_{q_{3},d}
+\delta_{q_{1},u}\delta_{q_{2},d}\delta_{q_{3},u}    \nonumber   \\
&&~~~~~~~~~~~~~~~~~+\delta_{q_{1},d}\delta_{q_{2},u}\delta_{q_{3},u} \bigr). 
\end{eqnarray} }We recall that, in the factorized case considered here, the flavor-dependent part 
$\mathcal {R}^{(f)}_{M_j,q_{1}\bar{q}_{2}}(N_{q_i},N_{\bar{q}_i})$ of the kernel function
represents the probability for the specified $q_1\bar q_2$ with the specified flavor $q_1$ and $\bar q_2$ from 
the system consisting of $N_q$ quarks
and $N_{\bar q}$ antiquarks to form the specified meson $M_j$ under
the condition that they are known to form a meson. Although
we can not prove it, it is very unlikely that this probability still
depends strongly on the environment. We therefore consider
the simplified case where $C_{M_j}$ is taken as a constant independent of $N_q$ or
$N_{\bar q}$. The same applies to $\mathcal {R}^{(f)}_{B_j,q_{1}q_{2}q_{3}}(N_{q_i},N_{\bar{q}_i})$.
In this case, we have
{\setlength\arraycolsep{0.2pt}
\begin{eqnarray}
 &&\overline{N}_{M_j}(N_{q_i},N_{\bar{q}_i})=C_{M_j} \frac{N_{q_1 \bar{q}_2}}{N_{q \bar{q}}}
 \overline{N}_{M}(N_{q},N_{\bar{q}},\sqrt{s}),         \\
 &&\overline{N}_{B_j}(N_{q_i},N_{\bar{q}_i})= N_{iter}C_{B_j}
 \frac{N_{q_1q_2q_3}}{N_{qqq}} \overline{N}_{B}(N_{q},N_{\bar{q}},\sqrt{s}), ~~~~~~~
\end{eqnarray} }where $N_{iter}$ stands for the number of possible iterations of $q_1q_2q_3$ which is $1$, $3$, and $6$
for three identical flavor, two different flavor, and three different flavor cases, respectively.

In the case when only $J^P=0^-$ and $1^-$ mesons and $J^P=\frac{1}{2}^+$ and $\frac{3}{2}^+$ baryons are considered,
we have, for mesons,
\begin{equation}
C_{M_j} =  \left\{
\begin{array}{ll}
{1}/{(1+R_{V/P})}~~~~~~~~   \textrm{for } J^P=0^-  \textrm{ mesons},  \\
{R_{V/P}}/{(1+R_{V/P})}~~~~         \textrm{for } J^P=1^-  \textrm{ mesons},
\end{array} \right.
\end{equation}
where $R_{V/P}$ represents the ratio of the $J^P=1^-$ vector mesons to the $J^P=0^-$ pseudoscalar mesons of the same flavor composition;
and for baryons,
\begin{equation}
C_{B_j} =  \left\{
\begin{array}{ll}
{R_{O/D}}/{(1+R_{O/D})}~~~~   \textrm{for } J^P=({1}/{2})^+  \textrm{ baryons},  \\
{1}/{(1+R_{O/D})}~~~~~~~~         \textrm{for } J^P=({3}/{2})^+  \textrm{ baryons},
\end{array} \right.
\end{equation}
except that $C_{\Lambda}=C_{\Sigma^0}={R_{O/D}}/{(1+2R_{O/D})}$, $C_{\Sigma^{*0}}={1}/{(1+2R_{O/D})}$,
and $C_{\Delta^{++}}=C_{\Delta^{-}}=C_{\Omega^{-}}=1$.
Here, $R_{O/D}$ stands for the ratio of the $J^P=(1/2)^+$ octet to the $J^P=(3/2)^+$ decuplet baryons of the same flavor composition.
The two parameters $R_{V/P}$ and $R_{O/D}$ can be determined by using the data from different high-energy 
reactions \cite{Anisovich1973NPB,FLShao2005PRC,RODJPG1995}.

\subsection{Modeling $P(N_{q_{i}},\langle N_{q_{i}}\rangle,\sqrt{s})$} \label{subsMBnonet}

We consider three flavors $u$, $d$, and $s$ of quarks and antiquarks.
Inside the system of $N_q$ quarks and $N_{\bar q}$ antiquarks,
we suppose that each quark can take flavor $u$, $d$, or $s$ with given probability $p_u$, $p_d$, or $p_s$ independent of the others.
In this case, the numbers of $u$, $d$, and $s$ quarks inside the system at a given $N_q$ obey the multinominal distribution, i.e.,
\begin{equation}
B(N_{q_{i}};N_{q})=\frac{N_{q}!}{N_{u}!N_{d}!N_{s}!}p_{u}^{N_u}p_{d}^{N_d}p_{s}^{N_s}\delta_{N_q,N_u+N_d+N_s},
\label{eq:Nqi}
\end{equation}
where $p_u=p_d=1/(2+\lambda_q)$, $p_s=\lambda_q/(2+\lambda_q)$,  and $\lambda_q$ is the effective strangeness suppression factor for quarks.
Similarly, for the antiquarks, we have
\begin{equation}
B(N_{\bar{q}_{i}};N_{\bar{q}})=\frac{N_{\bar{q}}!}{N_{\bar{u}}!N_{\bar{d}}!
N_{\bar{s}}!}
p_{\bar{u}}^{N_{\bar{u}}}p_{\bar{d}}^{N_{\bar{d}}}p_{\bar{s}}^{N_{\bar{s}}}
\delta_{N_{\bar{q}},N_{\bar{u}}+N_{\bar{d}}+N_{\bar{s}}},
\label{eq:Nbarqi}
\end{equation}
where $p_{\bar{u}}=p_{\bar{d}}=1/(2+\lambda)$, $p_{\bar s}=\lambda/(2+\lambda)$,  and $\lambda$ is the strangeness suppression factor
for antiquark production.

In general, in high-energy heavy-ion collisions, the system contains the contributions of the net quarks coming from the incident nuclei.
Hence the effective strangeness suppression factor $\lambda_q$ for the quarks is different from $\lambda$ for the antiquarks, which
do not have influence from the net quarks.
Here, we keep them as distinguished from each other so that we can apply the results to different cases.
Furthermore, we emphasize that the system considered corresponds to a quark-antiquark system produced in an $AA$
collision in a limited kinematic region. The system is supposed to be a small part of the whole quark-antiquark
system produced in the reaction so that the influence from the
global flavor compensation is considered to be negligible. The
global flavor compensation can have some influence on the
flavor correlation in hadron production. Such a case was for
example discussed in \cite{Liang1991PRD}
for baryon-antibaryon flavor
correlation in $e^+ e^−$ annihilation where the number of quarks
was of the order of tens and the average yield of baryons
in an event was less than one. It was found that, even in that
case, the global flavor compensation does have some effect on
the flavor correlation but the effect is not very large. Hence,
for simplicity and clarity, we neglect them in the discussion
here.

At given $N_q$ and $N_{\bar q}$, we average over the distribution of the numbers of quarks and/or antiquarks for different flavors.
It can easily be shown that, for $q_1\neq q_2\neq q_3$,
\begin{eqnarray}
&&\sum_{N_{q_i}} N_{q_1}B(N_{q_i};N_q)=N_q p_{q_1}, \\
&&\sum_{N_{q_i}} N_{q_1}N_{q_2}N_{q_3}B(N_{q_i};N_q)=N_{qqq} p_{q_1} p_{q_2} p_{q_3}, \\
&&\sum_{N_{q_i}} N_{q_1}(N_{q_1}-1)N_{q_2}B(N_{q_i};N_q)=N_{qqq} p_{q_1}^2 p_{q_2},
\end{eqnarray} and similarly for others, so we obtain
{\setlength\arraycolsep{0.2pt}
\begin{eqnarray}
 \overline{N}_{M_j}(N_{q},N_{\bar q},\sqrt{s})
 &=&C_{M_j}\ p_{q_{1}}p_{\bar{q}_{2}} \overline{N}_{M}(N_{q},N_{\bar q},\sqrt{s}),        \label{eq:Mj} \\
 \overline{N}_{B_j}(N_{q},N_{\bar q},\sqrt{s})
 &=&N_{iter}C_{B_j}\ p_{q_{1}}p_{q_{2}}p_{q_{3}} \overline{N}_{B}(N_{q},N_{\bar q},\sqrt{s}). ~~~~~~  \label{eq:Bj}
\end{eqnarray} }

For a subsystem of quarks and antiquarks in a given kinematic region in $AA$ collisions at given energy $\sqrt{s}$,
$\langle N_{q} \rangle$ and $\langle N_{\bar q} \rangle$ are fixed while $N_{q}$ and $N_{\bar q}$ follow
the distributions $P_q(N_{q};\langle N_{q} \rangle)$ and $P_{\bar q}(N_{\bar q};\langle N_{\bar q} \rangle)$, respectively.
Hence, we need to average over these distributions and obtain
{\setlength\arraycolsep{0.2pt}
\begin{eqnarray}
 \langle N_{M_j}\rangle(\langle N_{q}\rangle,\langle N_{\bar q}\rangle,\sqrt{s})
  &=&C_{M_j}\ p_{q_{1}}p_{\bar{q}_{2}}\langle N_{M}\rangle,
 \label{eq:NMj_aver}     \\
 \langle N_{B_j}\rangle(\langle N_{q}\rangle,\langle N_{\bar q}\rangle,\sqrt{s})
 &=&N_{iter}C_{B_j}\ p_{q_{1}}p_{q_{2}}p_{q_{3}}\langle N_{B}\rangle, ~~~~~~
 \label{eq:NBj_aver}
\end{eqnarray} }where $\langle N_{M}\rangle$ and $\langle N_{B}\rangle$ are functions of
$\langle N_{q}\rangle$, $\langle N_{\bar q}\rangle$ and $\sqrt{s}$
and stand for the average total number of the mesons and that of the baryons
produced in the combination process. They are given by
{\setlength\arraycolsep{0.5pt}
\begin{eqnarray}
 \langle N_{M}\rangle
 &=&\sum \limits_{N_qN_{\bar q}}
 P_q(N_{q};\langle N_{q} \rangle) P_{\bar q}(N_{\bar q};\langle N_{q} \rangle)
 \overline{N}_{M}(N_{q},N_{\bar q},\sqrt{s}), \nonumber\\
 \langle N_{B}\rangle
 &=&\sum \limits_{N_qN_{\bar q}}
 P_q(N_{q};\langle N_{q} \rangle) P_{\bar q}(N_{\bar q};\langle N_{q} \rangle)
 \overline{N}_{B}(N_{q},N_{\bar q},\sqrt{s}). ~~~~~~~\nonumber
\end{eqnarray} }

We see that, in this case, for the directly produced hadrons, the ratios of the yields of different mesons,
those of different baryons and those of the antibaryons separately are constants depending on
the parameters $\lambda$, $\lambda_q$, $R_{V/P}$ and $R_{O/D}$.
In general the effective strangeness suppression factor $\lambda_q$ for quarks contains the influence
from the net quark contributions and can be dependent on $\langle N_q\rangle$ and $\langle N_{\bar q}\rangle$.
This leads to a dependence on $\langle N_q\rangle$ and $\langle N_{\bar q}\rangle$ even for such kinds of hadron yield ratios.
In the case that the net quark contribution is negligible, we have $\lambda_q=\lambda$; 
these kinds of hadron yield ratios become constants independent of $\langle N_q\rangle$ and $\langle N_{\bar q}\rangle$.
This should be the case for a subsample in the central rapidity region of the bulk quark-antiquark system produced
in $AA$ collisions at very high energies such as those at the LHC.
In this case, we have also $\langle N_B\rangle=\langle N_{\bar B}\rangle$ and this, together with $p_{q_{i}}=p_{\bar{q_{i}}}$,
leads to $\langle N_{B_{j}}\rangle=\langle N_{\bar{B_{j}}}\rangle$.
These are predictions that can be checked at the LHC.

\subsection{Including the decay contributions}

Including the decay contributions, we calculate the yields of different hadrons in the final state.
We denote the decay contribution from a hadron $h_i$ to $h_j$ by $Br(h_i\to h_j)$ and obtain
\begin{equation}
\langle N_{h_j}^f\rangle = \langle N_{h_j}\rangle+ \sum_{i\not=j} Br(h_i\to h_j) \langle N_{h_i}\rangle,
\end{equation}
where we use the superscript $f$ to denote the results for the final hadrons to differentiate them from those 
for the directly produced hadrons. Here we consider only the influence from
the decay of the short-lived hadrons but do not consider the
influences from the final-state interactions of the hadrons.

The value of $Br(h_i\to h_j)$ can be obtained easily from the materials given by the Particle Data Group \cite{PDG2010}.
In the following, we take the strong and the electromagnetic decays into account.
For most of the hadrons, the results look very simple.
In the case in which only $J^P=0^-$ and $1^-$ mesons and $J^P=(1/2)^+$ and $(3/2)^+$ baryons are included,
the average yields of final hadrons, e.g., $K^{+}$, $p$, and $\Lambda$, are given as
{\setlength\arraycolsep{0.3pt}
\begin{eqnarray}
&&\langle N_{K^{+}}^{f}\rangle=\langle N_{K^{+}}\rangle
+\frac{2}{3}\langle N_{K^{*0}}\rangle +\frac{1}{3}\langle N_{K^{*+}}\rangle+0.489\langle N_{\phi}\rangle,~~~~  \\
&&\langle N_{p}^{f}\rangle=\langle N_{p}\rangle+\langle N_{\Delta^{++}}\rangle
+\frac{2}{3}\langle N_{\Delta^{+}}\rangle+\frac{1}{3}\langle N_{\Delta^{0}}\rangle, \\
&&\langle N_{\Lambda}^{f}\rangle=\langle N_{\Lambda}\rangle +\langle N_{\Sigma^{0}}\rangle
              \nonumber       \\
&&~~~~~~~~~~~~+0.883\langle N_{\Sigma^{*0}}\rangle+0.94(\langle N_{\Sigma^{*+}}\rangle+\langle N_{\Sigma^{*-}}\rangle).
\end{eqnarray} }

We consider the case discussed in Sec.~\ref{subsMBnonet} and substitute the results for $\langle N_{M_j}\rangle$ and those for
$\langle N_{B_j}\rangle$ given by Eqs.~(\ref{eq:NMj_aver}) and (\ref{eq:NBj_aver}) into the above equations 
and obtain 
{\setlength\arraycolsep{0.3pt}
\begin{eqnarray}
&&\langle N_{K^{+}}^{f}\rangle=p_up_{\bar s} \langle N_M\rangle
+\frac{0.489R_{V/P}}{1+R_{V/P}}p_sp_{\bar s} \langle N_M\rangle ,  \\
&&\langle N_{p}^{f}\rangle=4p_u^3\langle N_B\rangle,\\
&&\langle N_{\Lambda}^{f}\rangle=\left(\frac{5.30+12R_{O/D}}{2R_{O/D}+1}+\frac{5.64}{R_{O/D}+1}\right) p_{u}^{2}p_{s}\langle N_{B}\rangle,
\end{eqnarray} }where we have taken $p_u=p_d$ and $p_{\bar u}=p_{\bar d}$.

Taking $R_{V/P}=3$ according to the spin counting, 
but $R_{O/D}=2$ since decuplet baryon production is observed much more suppressed \cite{Anisovich1973NPB,FLShao2005PRC,RODJPG1995},
we then have
{\setlength\arraycolsep{0.3pt}
\begin{eqnarray}
&&\langle N_{K^{+}}^{f}\rangle=\bigl(p_up_{\bar s}+0.37p_sp_{\bar s}\bigr) \langle N_M\rangle ,  \\
&&\langle N_{\Lambda}^{f}\rangle=7.74p_{u}^{2}p_{s} \langle N_B\rangle.
\end{eqnarray} }

In Table~\ref{tab_yields}, we show the results obtained for baryons and antibaryons and those for strange mesons in different cases. 
We see that, like for protons, the results for many final baryons look even simpler than those for the directly produced cases
since the corresponding decuplet baryons decay strongly to these baryons.
This makes the average final yields for these baryons independent of the ratio $R_{O/D}$.
In Table~\ref{tab_yields}, we also present the results for the simple case without net quarks.
The results obtained if we take $R_{V/P}=3$ and $R_{O/D}=2$ are also given.

\begin{table*}[htbp]
\caption{Average yields of the directly produced hadrons and those including decay contributions. 
Here, in the second column, we show the results for the directly produced hadrons. 
In the third column, we show the results when the strong and the electromagnetic (S \& EM) decay contributions are taken into account.
The fourth column shows the results when the net quark influence is negligible. 
In the last two columns, we see the results for the case when $R_{V/P}=3$ and $R_{O/D}=2$.}
\renewcommand\arraystretch{1.7}
\begin{tabular}{@{}c|c|c|c|c|c@{}}
\toprule
       &   &    &   &\multicolumn{2}{c}{$R_{V/P}=3, R_{O/D}=2$} \\
\cline{5-6}
\raisebox{3.0ex}[0pt]{Hadron} &\raisebox{3.0ex}[0pt]{Directly produced} &\raisebox{3.0ex}[0pt]{With S \& EM decays} &\raisebox{3.0ex}[0pt]{$N_q^{net}= 0$ ($\lambda_q=\lambda$)} &$N_q^{net}\neq0$  &$N_q^{net}=0$ \\

\colrule
$p$            &$\frac{3R_{O/D}}{1+R_{O/D}} p_{u}^{3} \langle N_B\rangle$
                &$4p_{u}^{3} \langle N_B\rangle$
                 &$\frac{4}{(2+\lambda)^{3}}\langle N_B\rangle$
                  &$4p_{u}^{3} \langle N_B\rangle$
                   &$\frac{4}{(2+\lambda)^{3}}\langle N_B\rangle$\\

$n$            &$\frac{3R_{O/D}}{1+R_{O/D}} p_{u}^{3} \langle N_B\rangle$
                &$4p_{u}^{3} \langle N_B\rangle$
                 &$\frac{4}{(2+\lambda)^{3}}\langle N_B\rangle$
                  &$4p_{u}^{3} \langle N_B\rangle$
                   &$\frac{4}{(2+\lambda)^{3}}\langle N_B\rangle$ \\

$\Xi^0$        &$\frac{3R_{O/D}}{1+R_{O/D}} p_{u}p_{s}^{2} \langle N_B\rangle$
                &$3p_{u}p_{s}^{2} \langle N_B\rangle$
                 &$\frac{3\lambda^{2}}{(2+\lambda)^{3}}\langle N_B\rangle$
                  &$3p_{u}p_{s}^{2} \langle N_B\rangle$
                   &$\frac{3\lambda^{2}}{(2+\lambda)^{3}}\langle N_B\rangle$ \\

$\Xi^-$        &$\frac{3R_{O/D}}{1+R_{O/D}} p_{u}p_{s}^{2} \langle N_B\rangle$
                &$3p_{u}p_{s}^{2} \langle N_B\rangle$
                 &$\frac{3\lambda^{2}}{(2+\lambda)^{3}}\langle N_B\rangle$
                  &$3p_{u}p_{s}^{2} \langle N_B\rangle$
                   &$\frac{3\lambda^{2}}{(2+\lambda)^{3}}\langle N_B\rangle$ \\

$\Omega^-$     &$p_{s}^{3} \langle N_B\rangle$
                &$p_{s}^{3} \langle N_B\rangle$
                 &$\frac{\lambda^{3}}{(2+\lambda)^{3}}\langle N_B\rangle$
                  &$p_{s}^{3} \langle N_B\rangle$
                   &$\frac{\lambda^{3}}{(2+\lambda)^{3}}\langle N_B\rangle$\\

$\bar p$       &$\frac{3R_{O/D}}{1+R_{O/D}} p_{\bar u}^{3} \langle N_{\bar B}\rangle$
                &$4p_{\bar u}^{3} \langle N_{\bar B}\rangle$
                 &$\frac{4}{(2+\lambda)^{3}}\langle N_B\rangle$
                  &$4p_{\bar u}^{3} \langle N_{\bar B}\rangle$
                   &$\frac{4}{(2+\lambda)^{3}}\langle N_B\rangle$ \\

$\bar n$       &$\frac{3R_{O/D}}{1+R_{O/D}} p_{\bar u}^{3} \langle N_{\bar B}\rangle$
                &$4p_{\bar u}^{3} \langle N_{\bar B}\rangle$
                 &$\frac{4}{(2+\lambda)^{3}}\langle N_B\rangle$
                  &$4p_{\bar u}^{3} \langle N_{\bar B}\rangle$
                   &$\frac{4}{(2+\lambda)^{3}}\langle N_B\rangle$\\

$\bar \Xi^0$   &$\frac{3R_{O/D}}{1+R_{O/D}} p_{\bar u}p_{\bar s}^{2} \langle N_{\bar B}\rangle$
                &$3p_{\bar u}p_{\bar s}^{2} \langle N_{\bar B}\rangle$
                 &$\frac{3\lambda^{2}}{(2+\lambda)^{3}}\langle N_B\rangle$
                  &$3p_{\bar u}p_{\bar s}^{2} \langle N_{\bar B}\rangle$
                   &$\frac{3\lambda^{2}}{(2+\lambda)^{3}}\langle N_B\rangle$ \\

$\bar \Xi^+$   &$\frac{3R_{O/D}}{1+R_{O/D}} p_{\bar u}p_{\bar s}^{2} \langle N_{\bar B}\rangle$
                &$3p_{\bar u}p_{\bar s}^{2} \langle N_{\bar B}\rangle$
                 &$\frac{3\lambda^{2}}{(2+\lambda)^{3}}\langle N_B\rangle$
                  &$3p_{\bar u}p_{\bar s}^{2} \langle N_{\bar B}\rangle$
                   &$\frac{3\lambda^{2}}{(2+\lambda)^{3}}\langle N_B\rangle$ \\

$\bar\Omega^+$ &$p_{\bar s}^{3} \langle N_{\bar B}\rangle$
                &$p_{\bar s}^{3} \langle N_{\bar B}\rangle$
                 &$\frac{\lambda^{3}}{(2+\lambda)^{3}}\langle N_B\rangle$
                  &$p_{\bar s}^{3} \langle N_{\bar B}\rangle$
                   &$\frac{\lambda^{3}}{(2+\lambda)^{3}}\langle N_B\rangle$\\
\colrule

$K^{+}$        &$\frac{1}{1+R_{V/P}} p_up_{\bar s} \langle N_M\rangle$
                &$p_up_{\bar s}\bigl(1+\frac{0.49R_{V/P}}{1+R_{V/P}}\lambda_q\bigr) \langle N_M\rangle$
                 &$\frac{\lambda}{(2+\lambda)^{2}}\bigl(1+\frac{0.49R_{V/P}}{1+R_{V/P}}\lambda\bigr)
                 \langle N_M\rangle$
                  &$p_up_{\bar s}\bigl(1+0.37\lambda_q\bigr) \langle N_M\rangle$
                   &$\frac{\lambda+0.37\lambda^{2}}{(2+\lambda)^{2}}\langle N_M\rangle$\\

$K^{-}$        &$\frac{1}{1+R_{V/P}} p_{\bar u}p_s \langle N_M\rangle$
                &$p_{\bar u}p_s\bigl(1+\frac{0.49R_{V/P}}{1+R_{V/P}}\lambda\bigr) \langle N_M\rangle$
                 &$\frac{\lambda}{(2+\lambda)^{2}}\bigl(1+\frac{0.49R_{V/P}}{1+R_{V/P}}\lambda\bigr)
                 \langle N_M\rangle$
                  &$p_{\bar u}p_{s}\bigl(1+0.37\lambda\bigr) \langle N_M\rangle$
                   &$\frac{\lambda+0.37\lambda^{2}}{(2+\lambda)^{2}}\langle N_M\rangle$ \\

$K^{0}$        &$\frac{1}{1+R_{V/P}} p_up_{\bar s} \langle N_M\rangle$
                &$p_up_{\bar s}\bigl(1+\frac{0.34R_{V/P}}{1+R_{V/P}}\lambda_q\bigr) \langle N_M\rangle$
                 &$\frac{\lambda}{(2+\lambda)^{2}}\bigl(1+\frac{0.34R_{V/P}}{1+R_{V/P}}\lambda\bigr)
                 \langle N_M\rangle$
                  &$p_up_{\bar s}\bigl(1+0.26\lambda_q\bigr) \langle N_M\rangle$
                   &$\frac{\lambda+0.26\lambda^{2}}{(2+\lambda)^{2}}\langle N_M\rangle$\\

$\bar K^{0}$   &$\frac{1}{1+R_{V/P}} p_{\bar u}p_s \langle N_M\rangle$
                &$p_{\bar u}p_s\bigl (1+\frac{0.34R_{V/P}}{1+R_{V/P}}\lambda \bigr) \langle N_M\rangle$
                 &$\frac{\lambda}{(2+\lambda)^{2}}\bigl(1+\frac{0.34R_{V/P}}{1+R_{V/P}}\lambda\bigr)
                 \langle N_M\rangle$
                  &$p_{\bar u}p_s\bigl(1+0.26\lambda\bigr) \langle N_M\rangle$
                   &$\frac{\lambda+0.26\lambda^{2}}{(2+\lambda)^{2}}\langle N_M\rangle$\\

$\phi$        &$\frac{R_{V/P}}{1+R_{V/P}} p_sp_{\bar s} \langle N_M\rangle$
                &$\frac{R_{V/P}}{1+R_{V/P}} p_sp_{\bar s} \langle N_M\rangle$
                 &$\big(\frac{\lambda}{2+\lambda}\big)^{2}\frac{R_{V/P}}{1+R_{V/P}} \langle N_M\rangle$
                  &$\frac{3}{4} p_sp_{\bar s} \langle N_M\rangle$
                   &$\frac{3}{4}\big(\frac{\lambda}{2+\lambda}\big)^{2} \langle N_M\rangle$\\

$\Lambda$      &$\frac{6R_{O/D}}{1+2R_{O/D}} p_{u}^{2}p_{s} \langle N_{B}\rangle$
                &$\Bigl(\frac{5.30+12R_{O/D}}{1+2R_{O/D}}+\frac{5.64}{1+R_{O/D}}\Bigr)p_{u}^{2}p_{s} \langle N_B\rangle$
                 &$\frac{\lambda}{(2+\lambda)^{3}}\Bigl(\frac{5.30+12R_{O/D}}{1+2R_{O/D}}+\frac{5.64}{1+R_{O/D}}\Bigr) 
                  \langle N_B\rangle$
                  &$7.74p_{u}^{2}p_{s} \langle N_B\rangle$
                   &$\frac{7.74\lambda}{(2+\lambda)^{3}}\langle N_B\rangle$\\

$\Sigma^+$     &$\frac{3R_{O/D}}{1+R_{O/D}} p_{u}^{2}p_{s} \langle N_B\rangle$
                &$\Bigl(\frac{0.35}{1+2R_{O/D}}+\frac{0.18+3R_{O/D}}{1+R_{O/D}}\Bigr)p_{u}^{2}p_{s} \langle N_B\rangle$
                 &$\frac{\lambda}{(2+\lambda)^{3}}\Bigl(\frac{0.35}{1+2R_{O/D}}+\frac{0.18+3R_{O/D}}{1+R_{O/D}}\Bigr) \langle N_B\rangle$
                  &$2.13p_{u}^{2}p_{s} \langle N_B\rangle$
                   &$\frac{2.13\lambda}{(2+\lambda)^{3}}\langle N_B\rangle$\\

$\Sigma^0$     &$\frac{6R_{O/D}}{1+2R_{O/D}} p_{u}^{2}p_{s} \langle N_B\rangle$
                &$\Bigl(\frac{6R_{O/D}}{1+2R_{O/D}}+\frac{0.36}{1+R_{O/D}}\Bigr)p_{u}^{2}p_{s} \langle N_B\rangle$
                 &$\frac{\lambda}{(2+\lambda)^{3}}\Bigl(\frac{6R_{O/D}}{1+2R_{O/D}}+\frac{0.36}{1+R_{O/D}}\Bigr) \langle N_B\rangle$
                  &$2.52p_{u}^{2}p_{s} \langle N_B\rangle$
                   &$\frac{2.52\lambda}{(2+\lambda)^{3}}\langle N_B\rangle$\\

$\Sigma^-$     &$\frac{3R_{O/D}}{1+R_{O/D}} p_{u}^{2}p_{s} \langle N_B\rangle$
                &$\Bigl(\frac{0.35}{1+2R_{O/D}}+\frac{0.18+3R_{O/D}}{1+R_{O/D}}\Bigr)p_{u}^{2}p_{s} \langle N_B\rangle$
                 &$\frac{\lambda}{(2+\lambda)^{3}}\Bigl(\frac{0.35}{1+2R_{O/D}}+\frac{0.18+3R_{O/D}}{1+R_{O/D}}\Bigr) \langle N_B\rangle$
                  &$2.13p_{u}^{2}p_{s} \langle N_B\rangle$
                   &$\frac{2.13\lambda}{(2+\lambda)^{3}}\langle N_B\rangle$\\

$\bar\Lambda$  &$\frac{6R_{O/D}}{1+2R_{O/D}} p_{\bar u}^{2}p_{\bar s} \langle N_{\bar B}\rangle$
                &$\Bigl(\frac{5.30+12R_{O/D}}{1+2R_{O/D}}+\frac{5.64}{1+R_{O/D}}\Bigr)p_{\bar u}^{2}p_{\bar s} \langle N_{\bar B}\rangle$
                 &$\frac{\lambda}{(2+\lambda)^{3}}\Bigl(\frac{5.30+12R_{O/D}}{1+2R_{O/D}}+\frac{5.64}{1+R_{O/D}}\Bigr) 
                  \langle N_B\rangle$
                  &$7.74p_{\bar u}^{2}p_{\bar s} \langle N_{\bar B}\rangle$
                   &$\frac{7.74\lambda}{(2+\lambda)^{3}}\langle N_{B}\rangle$\\

$\bar\Sigma^-$ &$\frac{3R_{O/D}}{1+R_{O/D}} p_{\bar u}^{2}p_{\bar s} \langle N_{\bar B}\rangle$
                &$\Bigl(\frac{0.35}{1+2R_{O/D}}+\frac{0.18+3R_{O/D}}{1+R_{O/D}}\Bigr)p_{\bar u}^{2}p_{\bar s} \langle N_{\bar B}\rangle$
                 &$\frac{\lambda}{(2+\lambda)^{3}}\Bigl(\frac{0.35}{1+2R_{O/D}}+\frac{0.18+3R_{O/D}}{1+R_{O/D}}\Bigr) \langle N_B\rangle$
                  &$2.13p_{\bar u}^{2}p_{\bar s} \langle N_{\bar B}\rangle$
                   &$\frac{2.13\lambda}{(2+\lambda)^{3}}\langle N_{B}\rangle$\\

$\bar\Sigma^0$ &$\frac{6R_{O/D}}{1+2R_{O/D}} p_{\bar u}^{2}p_{\bar s} \langle N_{\bar B}\rangle$
                &$\Bigl(\frac{6R_{O/D}}{1+2R_{O/D}}+\frac{0.36}{1+R_{O/D}}\Bigr)p_{\bar u}^{2}p_{\bar s} \langle N_{\bar B}\rangle$
                 &$\frac{\lambda}{(2+\lambda)^{3}}\Bigl(\frac{6R_{O/D}}{1+2R_{O/D}}+\frac{0.36}{1+R_{O/D}}\Bigr) \langle N_B\rangle$
                  &$2.52p_{\bar u}^{2}p_{\bar s} \langle N_{\bar B}\rangle$
                   &$\frac{2.52\lambda}{(2+\lambda)^{3}}\langle N_{B}\rangle$\\

$\bar\Sigma^+$ &$\frac{3R_{O/D}}{1+R_{O/D}} p_{\bar u}^{2}p_{\bar s} \langle N_{\bar B}\rangle$
                &$\Bigl(\frac{0.35}{1+2R_{O/D}}+\frac{0.18+3R_{O/D}}{1+R_{O/D}}\Bigr)p_{\bar u}^{2}p_{\bar s} \langle N_{\bar B}\rangle$
                 &$\frac{\lambda}{(2+\lambda)^{3}}\Bigl(\frac{0.35}{1+2R_{O/D}}+\frac{0.18+3R_{O/D}}{1+R_{O/D}}\Bigr) \langle N_B\rangle$
                  &$2.13p_{\bar u}^{2}p_{\bar s} \langle N_{\bar B}\rangle$
                   &$\frac{2.13\lambda}{(2+\lambda)^{3}}\langle N_{B}\rangle$\\
\botrule
\end{tabular}
\label{tab_yields}
\end{table*}

We do not list the corresponding results for pions in Table~\ref{tab_yields} since the corresponding expressions are quite long. 
This is because the pion receives contributions from 
the decays of almost all the other mesons, baryons, and antibaryons. For example, for $\pi^+$, we have
\begin{widetext}
{\setlength\arraycolsep{0.3pt}
\begin{eqnarray}
\langle N_{\pi^{+}}^{f}\rangle=&&\langle N_{\pi^{+}}\rangle+\langle N_{\rho^{+}}\rangle
+\langle N_{\rho^{0}}\rangle+\frac{2}{3}(\langle N_{\bar{K}^{*0}}\rangle
+\langle N_{K^{*+}}\rangle)+0.2734\langle N_{\eta}\rangle
+0.9073\langle N_{\omega}\rangle+0.9274\langle N_{\eta'}\rangle
+0.1568\langle N_{\phi}\rangle
        \nonumber     \\
+&&0.94(\langle N_{\Sigma^{*+}}\rangle+\langle N_{\bar{\Sigma}^{*+}}\rangle)
+0.0585(\langle N_{\Sigma^{*0}}\rangle+\langle N_{\bar{\Sigma}^{*0}}\rangle)
+\frac{2}{3}(\langle N_{\Xi^{*0}}\rangle+\langle N_{\bar{\Xi}^{*+}}\rangle)
+\langle N_{\Delta^{++}}\rangle
+\frac{1}{3}(\langle N_{\Delta^{+}}\rangle+\langle N_{\bar \Delta^{0}}\rangle)
+\langle N_{\bar{\Delta}^{+}}\rangle. ~~~~~~~~
\label{eq:pi_decay}
\end{eqnarray} }By inserting the results given by Eqs.~(\ref{eq:NMj_aver}) and (\ref{eq:NBj_aver}), we obtain
{\setlength\arraycolsep{0.3pt} \begin{eqnarray}
\langle N_{\pi^{+}}^{f}\rangle
=&&\frac{1.71+2.91R_{V/P}}{1+R_{V/P}}p_{u}p_{\bar{u}}\langle N_{M}\rangle
+\frac{0.49+0.16R_{V/P}}{1+R_{V/P}}p_{s}p_{\bar{s}}\langle N_{M}\rangle
+\frac{2}{3}~\frac{R_{V/P}}{1+R_{V/P}}p_{\bar{u}}p_{s}\langle N_{M}\rangle
           \nonumber        \\
&&+\frac{2}{3}~\frac{R_{V/P}}{1+R_{V/P}}p_{u}p_{\bar{s}}\langle N_{M}\rangle
+(\frac{2.82}{1+R_{O/D}}+\frac{0.35}{1+2R_{O/D}})p_{u}^{2}p_{s}\langle N_{B}\rangle
+(\frac{2.82}{1+R_{O/D}}+\frac{0.35}{1+2R_{O/D}})p_{\bar{u}}^{2}p_{\bar{s}}\langle N_{\bar{B}}\rangle
           \nonumber        \\
&&+\frac{2}{1+R_{O/D}}p_{u}p_{s}^{2}\langle N_{B}\rangle
+\frac{2}{1+R_{O/D}}p_{\bar{u}}p_{\bar{s}}^{2}\langle N_{\bar{B}}\rangle
+\frac{2+R_{O/D}}{1+R_{O/D}}p_{u}^{3}\langle N_{B}\rangle
+\frac{2+R_{O/D}}{1+R_{O/D}}p_{\bar u}^{3}\langle N_{\bar B}\rangle.
\label{eq:pi_final}
\end{eqnarray} }For systems without net quarks, we have $p_{u}=p_{\bar{u}}=p_{d}=p_{\bar{d}}$ and $p_{s}=p_{\bar{s}}$, and thus we obtain
{\setlength\arraycolsep{0.3pt}
\begin{eqnarray}
\langle N_{\pi^{+}}^{f}\rangle =&&\frac{1.71+2.91R_{V/P}}{1+R_{V/P}}p_{u}^{2}\langle N_{M}\rangle
+\frac{0.49+0.16R_{V/P}}{1+R_{V/P}}p_{s}^{2}\langle N_{M}\rangle
+\frac{4}{3}～\frac{R_{V/P}}{1+R_{V/P}}p_{u}p_{s}\langle N_{M}\rangle
           \nonumber        \\
&&+(\frac{5.64}{1+R_{O/D}}+\frac{0.70}{1+2R_{O/D}})p_{u}^{2}p_{s}\langle N_{B}\rangle
+\frac{4}{1+R_{O/D}}p_{u}p_{s}^{2}\langle N_{B}\rangle
+\frac{4+2R_{O/D}}{1+R_{O/D}}p_{u}^{3}\langle N_{B}\rangle.
\end{eqnarray}  }If we take $R_{V/P}=3$ and $R_{O/D}=2$, we have
{\setlength\arraycolsep{0.3pt}
\begin{eqnarray}
\langle N_{\pi^{+}}^{f}\rangle = &&2.61p_{u}^{2}\langle N_{M}\rangle
+p_{u}p_{s}\langle N_{M}\rangle+0.24p_{s}^{2}\langle N_{M}\rangle
+\frac{4}{3}p_{u}p_{s}^{2}\langle N_{B}\rangle
+2.02p_{u}^{2}p_{s}\langle N_{B}\rangle
+\frac{8}{3}p_{u}^{3}\langle N_{B}\rangle
     \nonumber   \\
=&&\frac{2.61+\lambda+0.24\lambda^{2}}{(2+\lambda)^{2}}\langle N_{M}\rangle
+\frac{8/3+2.02\lambda+4/3\lambda^{2}}{(2+\lambda)^{3}}\langle N_{B}\rangle.~~~
\end{eqnarray}  }
\end{widetext}

From these results, we see clearly that there exist many simple relationships between the yields of different hadrons.
These are the characteristics for hadron production in the combination mechanism.
We will list some of these simple relations in the following.
Before doing that, we first discuss the net quark influences in the next section.

\subsection{Influence of the net quarks}

In a heavy-ion collision at high energy, the produced quark-antiquark system consisting of
the newly produced quarks and antiquarks and the net quarks from the incident nuclei.
For a subsample of this quark-antiquark system in a given kinematic region,
we have, in general,
\begin{equation}
 \langle N_{q}\rangle=\langle N_{\bar q}\rangle+\langle N_{q}^{net}\rangle .
\end{equation}
Both the momentum and flavor distributions of these net quarks are different from those for the newly produced ones, and
this leads to observable effects in the final hadrons produced in hadronization.
We expect that they have influences on the following aspects:

(i) The difference in momentum distribution leads to different $\gamma_M(N_q,N_{\bar q},\sqrt{s})$ and $\gamma_B(N_q,N_{\bar q},\sqrt{s})$,
as seen clearly from Eqs.~(\ref{eq:gammaM}) and (\ref{eq:gammaB}).
Furthermore, since $\langle N_q\rangle >\langle N_{\bar q}\rangle$, the average number of baryons, $\langle N_B\rangle$, should be
accordingly larger than $\langle N_{\bar B}\rangle$. The ratio $\langle N_B\rangle/\langle N_M\rangle$ should in general depend
on $\langle N_q^{net}\rangle/\langle N_q\rangle$.

(ii) The distribution of the number of quarks, $N_q$, at a given $\langle N_q\rangle$ is different from the corresponding distribution of the antiquarks.
The distribution of the number of the net quarks, $N_q^{net}$, at a given $\langle N_q^{net}\rangle$ is different from those
for the newly produced quarks and/or antiquarks. This leads to a difference between the quark number distribution $P_q(N_q,\langle N_q\rangle,\sqrt{s})$
and the antiquark number distribution $P_{\bar q}(N_{\bar q},\langle N_{\bar q}\rangle,\sqrt{s})$.

(iii) The flavor distribution of quarks is different from that for the antiquarks.
The net quarks take only two flavors, $u$ and $d$.
At a given $N_q^{net}$, the average numbers of $u$ and $d$ net quarks are determined by the numbers of protons and neutrons 
in the incident nuclei. For a given $AA$ collision,
$\overline N_u^{net}:\overline N_d^{net}=(A+Z):(2A-Z)$,
where $A$ and $Z$ are the numbers of nucleons and protons, respectively, in the incident nucleus $A$. The
numbers $N_u^{net}$ and $N_d^{net}$ follow a binominal distribution with $p_u^{net}:p_d^{net}=(A+Z):(2A-Z)$.
For the newly produced quarks or antiquarks, at a given $N_q^{new}$ or $N_{\bar q}$, the numbers of them of different flavors follow
the multinominal distribution as given by Eq.~(\ref{eq:Nqi}).
Hence, including the net quark contribution, the distribution of the numbers ($N_u$, $N_d$, and $N_s$) of the different flavors ($u$, $d$, and $s$)
of quarks at a given number of quarks ($N_q=N_u+N_d+N_s$) is also different from the corresponding distribution for the antiquarks.

The detailed calculations of the influences of these effects on the hadron yield ratios in the combination mechanism depend on the particular models.
In this paper, we present a rough estimate of, at least, the qualitative tendency of these effects by using the following two approximations.

First, we approximate that the flavor distribution of the number of quarks in the subsample of the system is independent of that for the antiquarks.
That for the antiquarks is given by the multinominal given by Eq.~(\ref{eq:Nbarqi}).
For the quarks, we approximate it by a multinominal distribution in the same form as that for the antiquarks but with different
probabilities of the different flavors. For the newly produced quarks, the flavor distribution should be $1:1:\lambda$ for $u$, $d$, and $s$ quarks,
the same as those for the antiquarks, but for the net quarks, it should be $(A+Z):(2A-Z):0$.
We take the average and obtain, at given $N_q$ and $N_q^{net}$,  
\begin{eqnarray}
 &&p_u=\frac{1}{2+\lambda}(1-\frac{N_q^{net}}{N_q})+\frac{A+Z}{3A} \frac{N_q^{net}}{N_q},  \\
 &&p_d=\frac{1}{2+\lambda}(1-\frac{N_q^{net}}{N_q})+\frac{2A-Z}{3A} \frac{N_q^{net}}{N_q},  \\
 &&p_s=\frac{\lambda}{2+\lambda}(1-\frac{N_q^{net}}{N_q}).
\end{eqnarray}
In the case in which the $u$ and $d$ difference is not very large, we neglect it and consider the case in which
$\overline N_u^{net}:\overline N_d^{net}=1:1$. In this case, we have, at given $N_q$ and $N_q^{net}$,
$p_u:p_d:p_s=1:1:\lambda_q$ and
\begin{equation}
\lambda_q=\frac{\overline N_s }{\overline N_u }
 =\lambda\Bigl[1+(1+\frac{\lambda}{2}) \frac{ N_q^{net} }{ N_q - N_q^{net} }\Bigr]^{-1}.  
 \label{eq:lambdaq}
\end{equation}
Under this approximation, we see that Eqs.~(\ref{eq:Mj}) and (\ref{eq:Bj}), the results we obtained in Sec.~\ref{subsMBnonet} for directly produced hadrons
in the combination of a quark-antiquark system at given $N_q$ and $N_{\bar{q}}$, are still valid.
We only need to note that the $p_{q_i}$ in this case is a function of $N_q$ and $N_q^{net}$ as given by Eq.~(\ref{eq:lambdaq}).
The influence of the isospin violation in net quarks can manifest
itself in the difference between the average yields of hadrons
belonging to the same charge multiplet. This can be studied
separately in experiments. Since our purpose is a rough estimation of the net quark influence, we consider in the following
first the simplified case where $u$ and $d$ are equal but leave the
isospin difference for future studies.

Second, for a subsample in a given kinematic region of the bulk system produced in a given $AA$ collision at given energy $\sqrt{s}$,
the averages $\langle N_{q} \rangle$ and  $\langle N_{\bar{q}} \rangle$ are fixed and the numbers $N_{q}$ and $N_{\bar{q}}$
follow the distribution $P(N_{q},N_{\bar{q}};\langle N_{q} \rangle,\langle N_{\bar{q}} \rangle,\sqrt{s})$. 
The average numbers of the hadrons produced should be the average over this distribution.
In general such averages depend on the precise form of $P(N_{q},N_{\bar{q}};\langle N_{q} \rangle,\langle N_{\bar{q}} \rangle,\sqrt{s})$.
In the rough estimations we made here, we approximate these averages by taking the corresponding values of the quantities
at the  averages $\langle N_{q} \rangle$ and $\langle N_{\bar{q}} \rangle$, i.e.,
\begin{equation}
 \langle N_{h_j}(N_q,N_{\bar q},\sqrt{s})\rangle \approx \overline{N}_{h_j} (\langle N_q\rangle, \langle N_{\bar q}\rangle,\sqrt{s}).
\end{equation}

Under these two approximations, all the results presented in the last four sections where we distinguish between $p_{q_i}$ and $p_{\bar q_i}$ apply
and we can use them to make estimates of the effects of the net quarks.

\subsection{Ratios of the yields of different hadrons}

From the results given in Table~\ref{tab_yields}, we see that there are many simple relations between the yields of different hadrons. 
In particular, for the directly produced hadrons, such relationships are very simple. 
Even for the final hadrons, although the decay influences are often very large, there still exists a set of 
simple relations between them.  
For example, independent of the values of $R_{V/P}$ and $R_{O/D}$, for the final hadrons where 
contributions from strong and electromagnetic decays are taken into account, we have  

{\setlength\arraycolsep{0.5pt}
\begin{eqnarray}
&&\frac{\langle N_{\Xi^{-}}^{f}\rangle}{\langle N_p^{f}\rangle}
=\frac{\langle N_{\Xi^{0}}^{f}\rangle}{\langle N_p^{f}\rangle}
=\frac{3}{4}\lambda_q^2,\\
&&\frac{\langle N_{\Omega^{-}}^{f}\rangle}{\langle N_p^{f}\rangle}
=\frac{1}{4}\lambda_q^3,\\
&&\frac{\langle N_{\bar\Xi^{+}}^{f}\rangle}{\langle N_{\bar p}^{f}\rangle}
=\frac{\langle N_{\bar\Xi^{0}}^{f}\rangle}{\langle N_{\bar p}^{f}\rangle}
=\frac{3}{4}\lambda^2,\\
&&\frac{\langle N_{\bar \Omega^{+}}^{f}\rangle}{\langle N_{\bar p}^{f}\rangle}
=\frac{1}{4}\lambda^3,\\
&&\frac{\langle N_{\bar{p}}^{f}\rangle}{\langle N_{p}^{f}\rangle}
=\Bigl(\frac{2+\lambda_q}{2+\lambda}\Bigr)^{3}\frac{\langle N_{\bar{B}}\rangle}{\langle N_{B}\rangle},      
\label{pbarp} \\ 
&&\frac{\langle N_{\bar{\Lambda}}^{f}\rangle}{\langle N_{\Lambda}^{f}\rangle}
=\Bigl(\frac{2+\lambda_q}{2+\lambda}\Bigr)^{3}\frac{\lambda}{\lambda_q}\frac{\langle N_{\bar{B}}\rangle}{\langle N_{B}\rangle},
                  \\ 
&&\frac{\langle N_{\bar{\Xi}^{+}}^{f}\rangle}{\langle N_{\Xi^{-}}^{f}\rangle}
=\Bigl(\frac{2+\lambda_q}{2+\lambda}\Bigr)^{3}\Bigl(\frac{\lambda}{\lambda_q}\Bigr)^{2}\frac{\langle N_{\bar{B}}\rangle}{\langle N_{B}\rangle},    \\
&&\frac{\langle N_{\bar{\Omega}^{+}}^{f}\rangle}{\langle N_{\Omega^{-}}^{f}\rangle}
=\Bigl(\frac{2+\lambda_q}{2+\lambda}\Bigr)^{3}\Bigl(\frac{\lambda}{\lambda_q}\Bigr)^{3}\frac{\langle N_{\bar{B}}\rangle}{\langle N_{B}\rangle}. \label{obaro}
\end{eqnarray} }

In the case in which net quark contribution is negligible, we have,  $\lambda_q=\lambda$ and $\langle N_B\rangle=\langle N_{\bar{B}}\rangle$, 
so that 
{\setlength\arraycolsep{0.5pt}
\begin{eqnarray}
&&\frac{\langle N_{\Xi^-}^{f}\rangle}{\langle N_p^{f}\rangle}
=\frac{\langle N_{\Xi^{0}}^{f}\rangle}{\langle N_p^{f}\rangle}
=\frac{3}{4}\lambda^2,\\
&&\frac{\langle N_{\Omega^{-}}^{f}\rangle}{\langle N_p^{f}\rangle}=
\frac{\langle N_{\bar \Omega^{+}}^{f}\rangle}{\langle N_{\bar p}^{f}\rangle}
=\frac{1}{4}\lambda^3, \label{obaro2}
\end{eqnarray}}and $\langle N_{\bar{B_j}}^{f}\rangle/\langle N_{B_j}^{f}\rangle=1$ for all the different types of $B_j$.

In the case in which $R_{V/P}=3$ and $R_{O/D}=2$, we have more such simple relations such as 
{\setlength\arraycolsep{0.5pt}
\begin{eqnarray}
&&\frac{\langle N_{K^{-}}^{f}\rangle}{\langle N_{K^{+}}^{f}\rangle}
=\Bigl(\frac{\lambda_q}{\lambda}\Bigr)\Bigl(\frac{1+0.37\lambda}{1+0.37\lambda_q}\Bigr), \label{kmkp} \\ 
&&\frac{\langle N_{\bar{\Lambda}}^{f}\rangle}{\langle N_{\bar{p}}^{f}\rangle}=1.935\lambda,
\label{lbpb} \\ 
&&\frac{\langle N_{\Lambda}^{f}\rangle}{\langle N_p^{f}\rangle}=1.935\lambda_q.
\label{lbpb}
\end{eqnarray}}These relations are intrinsic properties of the combination models in the sense that they do not depend on the details of particular combination models
but are determined mainly by the basic ideas of the combination mechanism.
They can be used to determine the free parameters and/or to test the mechanism.
We note in particular the following two features.

(i) For a system with equal average numbers of quarks and antiquarks, i.e., where net quark contributions are negligible,
these ratios are quite simple and can be tested by the data in extremely high energy collisions, e.g., at the LHC.

(ii) With net quark contributions, most of these yield ratios of different hadrons depend on $\langle N_q^{net}\rangle/\langle N_q\rangle$.
We see in particular that the ratios of the yields of hadrons to those of the corresponding antihadrons deviate from one in general.
They tend to one for reactions at very high energies where the net quark contribution tends to vanish.
This leads to an energy dependence of such ratios, even for particles such as $\Omega^-$ and $\bar\Omega^+$.
Such a property for the combination mechanism is different from what one expects from fragmentation and can be used as
a check to differentiate the different hadronization mechanisms.

We can also build some combinations of the average yields of the hadrons and obtain simple results for some more
sophisticated ratios 
such as,
{\setlength\arraycolsep{0.2pt}
\begin{eqnarray}
&&A\equiv \frac{~\langle N_{\bar{\Lambda}}\rangle \langle N_{K^{-}}\rangle \langle N_p\rangle~}
 {~\langle N_{\Lambda}\rangle \langle N_{K^{+}}\rangle \langle N_{\bar{p}}\rangle~}=1~,           
\label{eq:Adef}   \\
&&B\equiv \frac{~\langle N_{\Lambda} \rangle \langle N_{K^{-}}\rangle \langle N_{\bar{\Xi}^{+}}\rangle~}
 {~\langle N_{\bar{\Lambda}}\rangle \langle N_{K^{+}}\rangle \langle N_{\Xi^{-}}\rangle~}=1~,           
\label{eq:Bdef}   \\
&&d^{\Lambda}_{\Xi} \equiv \frac{~\langle N_{\bar{\Lambda}}\rangle
 \langle N_{\bar{\Xi}^{+}}\rangle~}{~\langle N_{\Lambda}\rangle \langle N_{\Xi^{-}}\rangle~}
 =\Bigl(\frac{2+\lambda_q}{2+\lambda}\Bigr)^6\Bigl(\frac{\lambda}{\lambda_q}\Bigr)^3
 \Bigl(\frac{\langle N_{\bar B}\rangle}{\langle N_B\rangle}\Bigr)^2~,  
\label{eq:dLambdadef}    \\
&&d^{p}_{\Omega} \equiv \frac{~\langle N_{\bar{p}}\rangle \langle N_{\bar{\Omega}^{+}}\rangle~}
{~\langle N_{p}\rangle \langle N_{\Omega^{-}}\rangle~}= d^{\Lambda}_{\Xi}~.
\label{eq:dXidef}
\end{eqnarray} }

We see that they all lead to simple results and these relations are not influenced by the resonance decays except 
$\phi \rightarrow K^{+}K^{-}$, which slightly changes $A$ and $B$ from unity. 
For the case in which $N_q^{net}=0$, all four ratios are equal to unity.

As a brief summary, we emphasize once more that the
model that we consider in this section is intended to be a general case based on the basic ideas of the combination 
mechanism. The purpose is to concentrate on the hadron yield correlations in the combination models. No effort is made to study
other properties such as momentum distribution, etc. The results obtained follow from the basic ideas of the 
combination mechanism and a number of assumptions, simplifications,
and/or approximations such as the factorization of flavor and
momentum dependence of the kernel functions, the flavor independence of the quark-antiquark momentum distribution, 
the independent production of different flavors of quarks
and antiquarks, and the approximations made in considering the net quark influences. These results do not depend
on the detailed form of the momentum dependence of kernel
functions and/or the momentum distributions of the quarks
and antiquarks. They even do not depend on whether
the quark number conservation or depletion is imposed in the
combination process. Such a conservation or depletion of
quark number influences the relationship between the average number of mesons (or baryons, or antibaryons) produced and
the number of quarks and/or antiquarks participating in the combination process (see in particular the discussion in 
\cite{ZWLinJPG}) but
does not influence the hadron yield ratios if the factorization
is assumed. These results should also be valid in the combination models discussed in the literature wherever these 
assumptions and/or approximations are also made (explicitly or implicitly).
Some of them should even be common in these
different models \cite{Zimanyi2000PLB,Greco2003PRL,Fries2003PRL,RCHwa2003PRC,FLShao2005PRC,FLShao2007PRC}. In fact, some of the relationships
presented above have also been derived in these literature \cite{Zimanyi2000PLB,Greco2003PRL,Fries2003PRL,RCHwa2003PRC,FLShao2005PRC,FLShao2007PRC}. For example, relations similar to those given
in Eqs.~(\ref{eq:Adef})-(\ref{eq:dXidef}) have been obtained in \cite{Zimanyi2000PLB}. These relations
can be used to test the validity of the combination mechanism
and the assumptions made. We will compare them with the
data available in the next section.

\section{Comparison with data}

There are already quite abundant data available from experiments in a quite broad energy region, from low SPS energies to RHIC and LHC energies \cite{20A30A2008PRC,plxoPRCSPS,plxoPRC62200,pPRC200,pPRC130,lxoPLB130,LPRL130,PLPRC2006,hypRL200,LPRL1588040,
MULSPRL130,OSPRL15840,PPLB62,kPRC4080158,pikp2011nuclex,sBpi2012nuclex}.
We compare the results obtained in the last section with these data in the
following.

As mentioned earlier, the study presented in Sec.~II is intended to be a general case for the combination of
a quark-antiquark system consisting of $N_q$ quarks and $N\bar q$
antiquarks. No effort is made to ascertain whether
the combination mechanism dominates the production of
hadrons in the given kinematic region in $AA$ collisions. In
this section, we choose the data in the central rapidity regions
in different $AA$ collisions at different energies. The agreement and/or disagreement of the theoretical results with these
data should give us a signature of whether the combination
mechanism with the above-mentioned assumptions and/or approximations is applicable. Also, since some of the theoretical
results depend on more inputs and some of them depend on
fewer inputs, we make the comparison at different levels.

\subsection{Comparison with LHC data}

At the first level, we consider a subsample of the quark-antiquark system in the central rapidity region produced 
in $AA$ collisions at very high energies. 
We suppose the energies are very high and the subsample that we consider is only a small part of the whole quark-antiquark system produced 
in the collision process so that the influence of the net quarks and that from the global flavor compensation are negligible. 
In this limiting case, the results for the ratios of the yields of different hadrons are divided into three classes.

In the first class, we consider the ratios of the yields of hadrons to those of the corresponding antihadrons.
Such ratios are unity, independent of any parameter.
This can be considered as a criterion for the validity of this limiting case.
Results from LHC experiments can be considered as an example for this case.
In the first three lines of Table~\ref{Tab:LHCpre}, we show the available experimental results for the particle to antiparticle ratios 
such as  $\langle N_{\pi^{-}}^{f}\rangle/\langle N_{\pi^{+}}^{f}\rangle$, $\langle N_{K^{-}}^{f}\rangle/\langle N_{K^{+}}^{f}\rangle$, 
and $\langle N_{\bar p}^{f}\rangle/\langle N_{p}^{f}\rangle$ 
at mid-rapidity.
The data are obtained from Ref. \cite{pikp2011nuclex}.
We see that they are indeed very close to unity.

\begin{table}[!htb]
\renewcommand{\arraystretch}{1.1}
\caption{Hadron yield ratios obtained for $N_q^{net}=0$ compared with 
data from the LHC in Pb + Pb collisions at $\sqrt{s}=2.76$ TeV.
The data are taken from Refs. \cite{pikp2011nuclex,sBpi2012nuclex}. The experimental result for $K^-/\pi^{-}$ is used to determine the strangeness 
suppression factor $\lambda$.}
\begin{tabular}{p{70pt}p{70pt}p{70pt}}
\toprule Ratios &Data & Calculations \\
\colrule
$\pi^{-}/\pi^{+}$   &$1.000\pm0.080$    &1 \\
$~K^{-}/K^{+}~$   &$0.987\pm0.076$    &1 \\
$\bar p/p$   &$0.995\pm0.077$    &1 \\
\colrule
$~\phi/K^{+}~$   &---   &0.278 \\
$~K_{S}^{0}/K^{+}~$   &---   &0.959 \\
$\Lambda/p$   &---    &0.832 \\
$\Xi^{-}/p$   &---    &0.139 \\
$\Omega^{-}/p$   &---    &0.020 \\
\colrule
$K^-/\pi^{-}$   &$0.155\pm0.012$    &0.155 \\
$p/\pi^{+}$   &$0.045\pm0.004$    &0.043 \\
$\Lambda/\pi^{+}$   &---    &0.036 \\
$\Xi^{-}/\pi^{+}$   &$0.005\pm0.001$    &0.006 \\
$\Omega^{-}/\pi^{+}$   &$~0.001\pm0.0002~$    &0.001 \\
$p/K^{+}$   &---    &0.275 \\
$\Lambda/K^{+}$   &---    &0.229 \\
$\Xi^{-}/K^{+}$   &---    &0.038 \\
$\Omega^{-}/K^{+}$   &---    &0.005 \\
\botrule
\end{tabular} \label{Tab:LHCpre}
\end{table}

In the second class, we consider the ratios of the average yields of hadrons such as $\phi/K^{+}$, $\Lambda/p$, and so on 
as shown in the second part of Table~\ref{Tab:LHCpre} (from the forth line to the eighth line). 
Because in this case we have $\langle N_B\rangle=\langle N_{\bar B}\rangle$ and $\lambda_q=\lambda$, 
these ratios depend only on one free parameter, the strangeness suppression factor $\lambda$, and are independent of the particular models.   
We can fix $\lambda$ by using the data for one ratio and make predictions for other particle ratios. 
Such results can be used to check the validity of the combination picture. 

In the third class, we consider the ratio of a specified meson to a specified baryon. 
To obtain the results for such ratios, we need the input for $\langle N_B\rangle/\langle N_M\rangle$. 
This can be slightly different in different combination models.
As an example, in the third part of Table~\ref{Tab:LHCpre} (from the ninth line to the end),  we show the results obtained by taking
\begin{equation}
\langle N_B \rangle/\langle N_M \rangle=1/12. 
\end{equation}
This is obtained by parametrizing the results for large $N_q=N_{\bar q}$ of the Monte Carlo generator (SDQCM) based on the combination rule developed 
by the Shandong group \cite{FLShao2005PRC,CEShao2009PRC,QBXie1988PRD} which has reproduced the data well.  
This parametrization is valid with high accuracy for $N_q$ larger than, say, 100. 
The parameter $\lambda$ is taken as $\lambda=0.43$ by fitting the data of $K^{-}/\pi^{-}$. 
We see that the theoretical results  agree with the LHC data whenever available.

\subsection{Comparison with RHIC and SPS data}

At the second level of comparison, we consider the case where $N_q^{net}\not=0$.
In this case, we need a further input of $\langle N_q^{net}\rangle/\langle N_q\rangle$ which describes the strength of the net quark influence.
Clearly, this ratio depends on the type and the energy of the incident nuclei, and also on the kinematic region that we consider. 
For example, to make a good comparison with the data from the RHIC and the
SPS \cite{20A30A2008PRC,plxoPRCSPS,plxoPRC62200,pPRC200,pPRC130,lxoPLB130,LPRL130,PLPRC2006,hypRL200,LPRL1588040,
MULSPRL130,OSPRL15840,PPLB62,kPRC4080158}, 
we need to take such effects into account.

In practice, to carry out the calculations, we first fix $\lambda_q$ and $\lambda$ by using the data for 
$\langle N_{\Xi^{-}}\rangle/\langle N_p\rangle$ to determine $\lambda_q$ and $\langle N_{\bar \Xi^{+}}\rangle/\langle N_{\bar p}\rangle$ to determine $\lambda$, 
and we then derive $\langle N_q^{net}\rangle/\langle N_q\rangle$ using Eq.~(\ref{eq:lambdaq}). 
In Table~\ref{tab:lamlamq}, we show the results obtained by taking the data for $dN/dy$ at $y=0$ (where $y$ denotes the rapidity of the hadron).
These results can be used to calculate the hadron ratios in the mid-rapidity regions. 

\begin{table}[!htb]
\renewcommand{\arraystretch}{1.4}
\caption{The fixed $\lambda$ and $\lambda_q$ and the derived $\langle N_q^{net}\rangle/\langle N_q\rangle$ at RHIC and SPS energies.}
\begin{tabular}{@{}c|c|c|c|c|c|c|c@{}}
\toprule
    &\multicolumn{3}{c|}{RHIC$\sqrt{s}$ (GeV)}  &\multicolumn{4}{c}{SPS $E_{beam}$ ($A$ GeV)} \\
\cline{2-8}
\raisebox{2.0ex}[0pt]{Energy} &200  &130  &62.4  &158  &80  &40  &30   \\
\colrule
$\lambda$   &0.425   &0.412   &0.369   &0.499   &0.567   &0.540   &0.645\\
$\lambda_q$   &0.397   &0.375   &0.328   &0.255   &0.232   &0.193   &0.192 \\
$\langle N_q^{net}\rangle/\langle N_q\rangle$   &0.056   &0.074   &0.095   &0.434   &0.529   &0.587   &0.640 \\
\botrule
\end{tabular}
\label{tab:lamlamq}
\end{table}

From the table, we see that $\langle N_q^{net}\rangle/\langle N_q\rangle$ is small at RHIC energies but quite 
large at SPS energies and the expected effects should be large at those energies.

Using the obtained $\lambda$ and $\lambda_q$ values as inputs, we calculate the hadron yield ratios that 
are independent of $\langle N_B\rangle/\langle N_M\rangle$.
The results are given in Table~\ref{Tab:RHICSPSpre1}. The corresponding data are from 
Refs. \cite{20A30A2008PRC,pPRC130,kPRC4080158,pPRC200,plxoPRCSPS,plxoPRC62200,
LPRL130,PLPRC2006,hypRL200,LPRL1588040,MULSPRL130,OSPRL15840,PPLB62}.

\begin{table*}[!htb]\footnotesize
\renewcommand{\arraystretch}{2.0}
\caption{Comparison of the calculated results for hadron yield ratios that are independent of $\langle N_B\rangle/\langle N_M\rangle$ with data from RHIC and SPS energies \cite{20A30A2008PRC,pPRC130,kPRC4080158,pPRC200,plxoPRCSPS,plxoPRC62200,LPRL130,PLPRC2006,
hypRL200,LPRL1588040,MULSPRL130,OSPRL15840,PPLB62}. The experimental result for ${\Xi^{-}}/{p}$ and that for $\bar \Xi^{+}/\bar p$ are used to
determine the strangeness suppression factor $\lambda_q$ and $\lambda$, respectively.}
\begin{tabular}{@{}c|cc|cc|cc|cc|cc|cc|cc@{}}
\toprule
       &\multicolumn{6}{c|}{$\sqrt{s}$ (GeV) for Au + Au at RHIC}   &\multicolumn{8}{c}{$E_{beam}$ ($A$ GeV) for Pb + Pb at SPS} \\
\cline{2-15}
\raisebox{0.2ex}[0pt]{Ratio} &\multicolumn{2}{c|}{200}  &\multicolumn{2}{c|}{130}  &\multicolumn{2}{c|}{62.4}  
&\multicolumn{2}{c|}{158}  &\multicolumn{2}{c|}{80}  &\multicolumn{2}{c|}{40}  &\multicolumn{2}{c}{30}   \\
\cline{2-15}
       &data &theory   &data &theory   &data &theory   &data &theory   &data &theory   &data &theory   &data &theory \\
\colrule
$\bar \Lambda/\bar p$    &$ 0.94\pm0.15 $ &0.82   &$ 0.93\pm0.34 $ &0.80   &$ 0.82\pm0.18  $ &0.71
&$0.98\pm0.23 $ &0.97   &$ 1.22\pm0.30 $ &1.10   &$ 1.31\pm0.35 $ &1.05   &$ 1.31\pm0.41 $ &1.25 \\
$\bar \Xi^{+}/\bar p$    &$ 0.14\pm0.03 $ &0.14   &$ 0.13\pm0.03 $ &0.13   &$ 0.10\pm0.03 $ &0.10
&$ 0.19\pm0.05 $ &0.19   &$ 0.24\pm0.07 $ &0.24   &$ 0.22\pm0.07 $ &0.22   &$ 0.31\pm0.15 $ &0.31 \\
$\bar \Omega^{+}/\bar p$    &------ &0.019   &------ &0.017   &$ 0.017\pm0.005 $ &0.013
&$ 0.042\pm0.020 $ &0.031   &------ &0.046   &------ &0.039   &------ &0.067 \\
\colrule
${\Lambda}/{p}$    &$ 0.91\pm0.15 $ &0.77   &$ 0.90\pm0.30 $ &0.73   &$ 0.78\pm0.16 $ &0.63
&$ 0.37\pm0.09 $ &0.49   &$ 0.45\pm0.08 $ &0.45   &$ 0.37\pm0.06 $ &0.37   &$ 0.35\pm0.06 $ &0.37 \\
${\Xi^{-}}/{p}$    &$ 0.12\pm0.02 $ &0.12   &$ 0.11\pm0.03 $ &0.11   &$ 0.08\pm0.02 $ &0.08
&$ 0.05\pm0.01 $ &0.05   &$ 0.04\pm0.01 $ &0.04   &$ 0.03\pm0.01 $ &0.03   &$ 0.03\pm0.01 $ &0.03 \\
${\Omega^{-}}/{p}$    &------ &0.016   &------ &0.013   &$ 0.010\pm0.003 $ &0.009
&$ 0.005\pm0.001 $ &0.004   &------ &0.003   &------ &0.002   &------ &0.002 \\
\colrule
${K^-}/{K^{+}}$    &$ 0.96\pm0.17 $ &0.94   &$ 0.92\pm0.09 $ &0.92   &$ 0.86\pm0.09 $ &0.90
&$ 0.57\pm0.05 $ &0.55   &$ 0.48\pm0.04 $ &0.46   &$0.38\pm0.04 $ &0.40   &$ 0.37\pm0.04 $ &0.34 \\
${\bar K^0}/{K^{0}}$    &------ &0.94   &------ &0.92   &------ &0.90
&------ &0.54   &------ &0.44   &------ &0.39   &------ &0.33 \\
${K^0}/{K^{+}}$    &------ &0.96   &------ &0.96   &------ &0.97
&------ &0.97   &------ &0.98   &------ &0.98   &------ &0.98 \\
\botrule
\end{tabular}
\label{Tab:RHICSPSpre1}
\end{table*}

We see in particular that the ratio of the yield of hadrons to that of the corresponding antihadrons is not unity in this case. 
We also note that the ratio such as $\langle N_{K^-}^f\rangle/\langle N_{K^+}^f\rangle=({\lambda_q}/{\lambda})(1+0.37\lambda)/(1+0.37\lambda_q)$ is a good 
example to show the change of the effective strange suppression for quarks.
The results should decrease monotonically with increasing $\langle N_q^{net}\rangle/\langle N_q\rangle$.   
There are data available for $\langle N_{K^-}^f\rangle/\langle N_{K^+}^f\rangle$ 
at different energies \cite{20A30A2008PRC,pPRC130,kPRC4080158} and the data show clearly that 
the ratio increases with increasing energy.
This qualitative tendency is consistent with the effect of net quark contribution since the relative influence of the net quarks
becomes smaller at higher energies.

To calculate other ratios, we need $\langle N_B\rangle/\langle N_M\rangle$ and $\langle N_{\bar B}\rangle/\langle N_M\rangle$.
For this purpose, we parametrize $\langle N_{\bar B}\rangle/\langle N_M\rangle$ for the case where $N_q^{net}\not =0$ 
using SDQCM \cite{FLShao2005PRC,CEShao2009PRC,QBXie1988PRD},  and we obtain
\begin{equation}
\frac{\langle N_{\bar B}\rangle}{\langle N_M\rangle}=\frac{1}{12}\Bigl(1-\frac{\langle N_q^{net}\rangle}{\langle N_q\rangle}\Bigr)^{2.8}. 
\label{eq:BbarM}
\end{equation}
We found that this parametrization is a good approximation to the results obtained from SDQCM for different $N_q$ in the range of $N_q>100$. 
In this model, quark number conservation or depletion is imposed in the combination process so that
$3\langle N_B\rangle+\langle N_M\rangle=\langle N_q\rangle$ and 
$3\langle N_{\bar B}\rangle+\langle N_M\rangle=\langle N_q\rangle-\langle N_q^{net}\rangle$,
and we derive
\begin{equation}
\frac{\langle N_{B}\rangle}{\langle N_M\rangle}=\Bigl(\frac{\langle N_{\bar B}\rangle}{\langle N_M\rangle}
+\frac{1}{3}\frac{\langle N_q^{net}\rangle}{\langle N_q\rangle}\Bigr)/\Bigl(1-\frac{\langle N_q^{net}\rangle}{\langle N_q\rangle}\Bigr).
\label{eq:BM}
\end{equation}
With Eqs.~(\ref{eq:BbarM}) and (\ref{eq:BM}), we calculate the ratios of the yields of different baryons to mesons. 
The results are shown in Table~\ref{Tab:RHICSPSpre2}.
The data are calculated from $dN/dy$ at $y=0$ for different energies from 
Refs. \cite{20A30A2008PRC,pPRC130,kPRC4080158,pPRC200,plxoPRCSPS,plxoPRC62200,
LPRL130,PLPRC2006,hypRL200,LPRL1588040,MULSPRL130,OSPRL15840,PPLB62}.

\begin{table*}[!htb]\footnotesize
\renewcommand{\arraystretch}{2.0}
\caption{Comparison of the calculated results for the baryon-to-meson ratios that are dependent of $\langle N_B\rangle/\langle N_M\rangle$ 
with data from RHIC and SPS 
energies \cite{20A30A2008PRC,pPRC130,kPRC4080158,pPRC200,plxoPRCSPS,plxoPRC62200,LPRL130,
PLPRC2006,hypRL200,LPRL1588040,MULSPRL130,OSPRL15840,PPLB62}.}
\begin{tabular}{@{}c|cc|cc|cc|cc|cc|cc|cc@{}}
\toprule
       &\multicolumn{6}{c|}{$\sqrt{s}$ (GeV) for Au + Au at RHIC}  &\multicolumn{8}{c}{$E_{beam}$ ($A$ GeV) for Pb + Pb at SPS} \\
\cline{2-15}
\raisebox{0.2ex}[0pt]{Ratio} &\multicolumn{2}{c|}{200}  &\multicolumn{2}{c|}{130}  &\multicolumn{2}{c|}{62.4}  
&\multicolumn{2}{c|}{158}  &\multicolumn{2}{c|}{80}  &\multicolumn{2}{c|}{40}  &\multicolumn{2}{c}{30}   \\
\cline{2-15}
       &data &theory   &data &theory   &data &theory   &data &theory   &data &theory   &data &theory   &data &theory \\
\colrule
${\bar p}/{K^{-}}$    &$ 0.27\pm0.05  $ &0.25    &$ 0.32\pm0.07 $ &0.25    &$ 0.31\pm0.05 $ &0.28
&$ 0.10\pm0.02 $ &0.08    &$ 0.07\pm0.01 $ &0.05    &$ 0.04\pm0.01 $ &0.04    &$ 0.02\pm0.01  $ &0.03  \\
${\bar \Lambda}/{K^{-}}$    &$ 0.26\pm0.04  $ &0.21    &$ 0.30\pm0.09 $ &0.20    &$ 0.26\pm0.04 $ &0.20
&$ 0.10\pm0.01 $ &0.08    &$ 0.09\pm0.02 $ &0.05    &$ 0.06\pm0.01 $ &0.04    &$ 0.03\pm0.01  $ &0.03  \\
${\bar \Xi^{+}}/{K^{-}}$    &$ 0.04\pm0.01  $ &0.03    &$ 0.04\pm 0.01$ &0.03    &$ 0.03\pm 0.01$ &0.03
&$ 0.018\pm 0.004$ &0.015    &$ 0.018\pm 0.004$ &0.012    &$ 0.009\pm0.003 $ &0.009    &$ 0.006\pm0.003  $ &0.008  \\
${\bar \Omega^{+}}/{K^{-}}$    &------ &0.005    &------ &0.004    &$ 0.005\pm 0.001$ &0.004
&$ 0.004\pm0.002 $ &0.003    &------ &0.002    &------ &0.002    &------ &0.002  \\
\colrule
${p}/{K^{+}}$    &$ 0.36\pm0.07  $ &0.33    &$0.42 \pm0.09 $ &0.36    &$0.54 \pm 0.08$ &0.44
&$ 1.00\pm0.15 $ &1.03    &$ 1.22\pm0.18 $ &1.32    &$ 2.05\pm 0.29$ &1.79    &$ 1.99\pm 0.36 $ &1.93  \\
${\Lambda}/{K^{+}}$    &$0.33 \pm 0.05 $ &0.25    &$ 0.37\pm 0.10$ &0.26    &$0.42\pm0.08$ &0.28 
&$0.37\pm0.08$ &0.51    &$ 0.55\pm 0.08$ &0.60    &$0.76\pm 0.09$ &0.67    &$0.69\pm0.10 $ &0.72  \\
${\Xi^{-}}/{K^{+}}$    &$ 0.04\pm0.01  $ &0.04    &$ 0.04\pm 0.01$ &0.04    &$ 0.04\pm0.01 $ &0.04
&$ 0.049\pm0.009 $ &0.050    &$ 0.050\pm 0.011$ &0.054    &$ 0.057\pm 0.013$ &0.050    &$ 0.055\pm0.014  $ &0.054  \\
${\Omega^{-}}/{K^{+}}$    &------ &0.005    &------ &0.005    &$ 0.006\pm0.001 $ &0.004 
&$ 0.005\pm0.001 $ &0.004    &------ &0.004    &------ &0.003    &------ &0.003  \\
\botrule
\end{tabular}
\label{Tab:RHICSPSpre2}
\end{table*}

\begin{figure}[htbp]
\centering
 \includegraphics[width=0.9\linewidth]{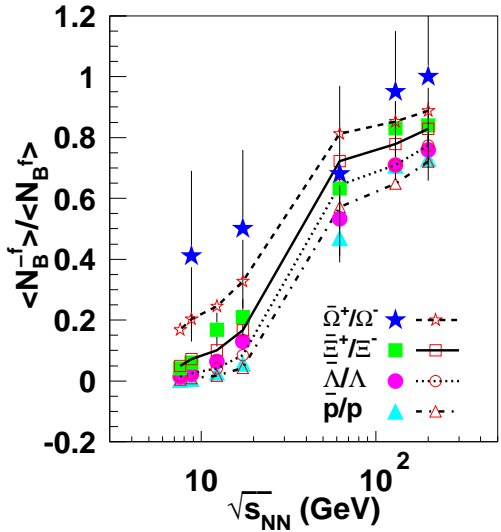}\\
 \caption{(Color online)The yield ratios of antibaryons to baryons at mid-rapidity at different energies. 
 The filled symbols with error bars are the experimental results taken from
 Refs. \cite{plxoPRCSPS,plxoPRC62200,pPRC200,pPRC130,lxoPLB130}.
 The corresponding open symbols represent the calculated results where the lines that 
 connect these symbols are just used to guide the eyes. }\label{BbarBratio}
\end{figure}

With the values of $\lambda$ and $\lambda_q$ given in Table~\ref{tab:lamlamq} and $\langle N_{\bar B}\rangle/\langle N_B\rangle$ 
derived from Eqs.~(\ref{eq:BbarM}) and (\ref{eq:BM}),  
we calculate in particular the ratios of the antibaryons to the corresponding baryons at different energies. 
The results obtained are shown in Fig.\,\ref{BbarBratio} with the open symbols connected by different lines to guide the eye. 
The filled symbols with error bars are the experimental
data taken from Refs. \cite{plxoPRCSPS,plxoPRC62200,pPRC200,pPRC130,lxoPLB130}.
From Fig.\,\ref{BbarBratio}, we see in particular that 
${\langle N_{\bar{\Omega}^{+}}^{f}\rangle}/{\langle N_{\Omega^{-}}^{f}\rangle}>
{\langle N_{\bar{\Xi}^{+}}^{f}\rangle}/{\langle N_{\Xi^{-}}^{f}\rangle}>
{\langle N_{\bar{\Lambda}}^{f}\rangle}/{\langle N_{\Lambda}^{f}\rangle}>
{\langle N_{\bar{p}}^{f}\rangle}/{\langle N_{p}^{f}\rangle}$. 
This was first pointed out by the NA49 Collaboration and was regarded as a distinct hierarchy of the antibaryon to baryon 
ratios \cite{plxoPRCSPS}.
The hierarchy can be naturally reproduced by Eqs.~(\ref{pbarp})-(\ref{obaro}) in the simple quark combination models.
We also see that the antibaryon to baryon ratios increase with increasing energy, indicating that the 
net quark influence becomes smaller at higher energies. 

It was considered as a surprise that the net quark also influences 
$\langle N_{\bar{\Omega}^{+}}^{f}\rangle/\langle N_{\Omega^{-}}^{f}\rangle$ significantly, 
as shown in Fig.~\ref{BbarBratio}, although the $\Omega$ hyperon does not consist of $u$ or $d$ quarks.  
We see that from the lowest SPS to the highest RHIC energy,  
$\langle N_{\bar{\Omega}^{+}}^{f}\rangle/\langle N_{\Omega^{-}}^{f}\rangle$ 
increases from about $0.4$ to unity. 
This cannot be understood in the fragmentation models but can be naturally explained 
in the framework of quark combination models. 
Because in the combination models more net quarks imply more chances for the antistrange quarks 
to meet quarks to form mesons, there will be 
more antistrange quarks exhausted to form kaons than strange quarks.
The probability for a strange antiquark to
combine with other antiquarks to form an antibaryon is
smaller than that for a strange quark to combine with other quarks to
form a baryon.
This leads to fewer $\bar \Omega$'s than $\Omega$'s for the same number of 
strange quarks and antistrange quarks.    
This qualitative tendency is consistent with data, and from the figure we see also the 
quantitative results agree well with the data.

\begin{figure}[htbp]
  \includegraphics[width=1.0\linewidth]{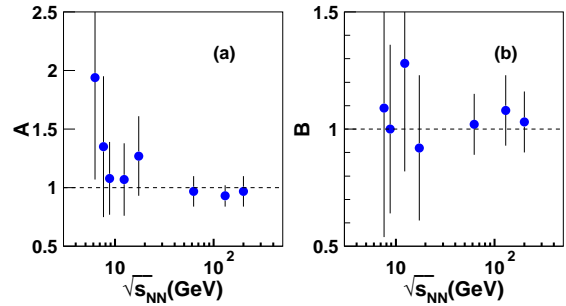}\\
  \caption{(Color online)The correlation quantities $A$ (a) and $B$ (b) as the function of the collision energy.
  The experimental data, filled circles with total statistical and systematic errors, are from 
  Refs. \cite{20A30A2008PRC,plxoPRCSPS,plxoPRC62200,pPRC200,pPRC130,lxoPLB130,kPRC4080158}.}
  \label{ab}
\end{figure}

We also compare the experimental results with the predictions shown in Eqs.~(\ref{eq:Adef})-(\ref{eq:dXidef}) 
for the ``more sophisticated" ratios $A=B=1$ and $d^{\Lambda}_{\Xi}=d^{p}_{\Omega}$.  
From the data available \cite{plxoPRCSPS,plxoPRC62200,pPRC200,pPRC130,lxoPLB130}, 
we calculate these ratios and show the results in Table~\ref{d-values} and Fig.\,\ref{ab} 
for $d^{\Lambda}_{\Xi}$ and $d^{p}_{\Omega}$ and $A$ and $B$, respectively. 
We see that the data are consistent with $d^{\Lambda}_{\Xi}=d^{p}_{\Omega}$. 
The results for $A$ and $B$ at RHIC energies are consistent with unity
while the error bars for those at SPS energies are too large to make a judgment.

\begin{table}[ht]
\renewcommand{\arraystretch}{1.2}
\caption{The deduced values of $d_{\Xi}^{\Lambda}$ and $d_{\Omega}^{p}$ at
RHIC and SPS energies. Errors shown are total errors. The data are from Refs \cite{plxoPRCSPS,plxoPRC62200,pPRC200,pPRC130,lxoPLB130}.}
\begin{tabular}{p{70pt}p{70pt}p{70pt}}
\toprule Energy &$d_{\Xi}^{\Lambda}$ &$d_{\Omega}^{p}$ \\
\colrule
200 GeV   &$0.64\pm0.07$    &$0.73\pm0.23$ \\
130 GeV   &$0.59\pm0.08$    &$0.67\pm0.15$ \\
62.4 GeV  &$0.34\pm0.04$    &$0.32\pm0.14$ \\
$158A$ GeV  &$0.027\pm0.009$  &$0.029\pm0.016$ \\
$40A$ GeV  &$0.0013\pm0.0005$  &$0.0032\pm0.0023$   \\
\botrule
\end{tabular} \label{d-values}
\end{table}

\section{summary}

We study the hadron yield correlations in the combination
models. With the basic ideas of the combination mechanism
and a few simplifications and/or assumptions based on symmetry and general principles, we show that the hadron yield ratios
can be calculated and have a series of regular properties. These
ratios are properties of the combination mechanism under these
assumptions and/or approximations such as the factorization of
flavor and momentum dependence of the kernel function, the
flavor independence of the momentum distributions, and the
approximations made for the net quark contributions. They are
independent of the particular models where usually particular
assumptions are made for the kernel functions. A systematic
study of these ratios should provide good hints as to whether
the combination mechanism is at work. Comparisons with
available data are made and predictions for future experiments
are given.

\section*{Acknowledgements}
The authors thank Zong-guo Si, Shi-yuan Li and other members of the particle theory group of Shandong University for helpful discussions.
This work is supported in part by the National Natural Science Foundation of China under Grants No. 11175104, No. 10947007, and No. 10975092 and
by the Natural Science Foundation of Shandong Province, China, under Grant No. ZR2011AM006.


\begin{thebibliography}{00}
\bibitem{Anisovich1973NPB}
V. V. Anisovich and V. M. Shekhter, Nucl. Phys. B {\bf55}, 455 (1973).
\bibitem{Bjorken1974PRD}
J. D. Bjorken and G. R. Farrar, Phys. Rev. D {\bf9}, 1449 (1974).
\bibitem{sdpi1981PRD}
A. Suzuki {\it et al.}, Phys. Rev. D {\bf24}, 605 (1981).
\bibitem{1982PRD}
Chandra Gupt, R. K. Shivpuri, N. S. Verma, and A. P. Sharma, Phys. Rev. D {\bf26}, 2202 (1982).
\bibitem{Liang1991PRD}
Z. T. Liang and Q. B. Xie, Phys. Rev. D {\bf43}, 751 (1991);
Q. Wang, Z. G. Si, and Q. B. Xie, Int. J. Mod. Phys. A {\bf11}, 5203 (1996).
\bibitem{thermal1995PLB}
P. Braun-Munzinger, J. Stachel, J. P. Wessels, and N. Xu, Phys.\ Lett.\ B {\bf344}, 43(1995).
\bibitem{Sa1995PRC}
B. H. Sa, A. Tai, and Z. D. Lu, Phys. Rev. C {\bf52}, 2069 (1995).
\bibitem{Bialas1998PLB}
A. Bialas, Phys.\ Lett.\ B {\bf442}, 449(1998).
\bibitem{Zimanyi2000PLB}
J. Zim\'anyi, T. S. Bir\'o, T. Cs\"org\H{o}, and P. L\'evai, Phys. Lett.  B {\bf472}, 243
(2000).
\bibitem{Greco2003PRL}
V. Greco, C. M. Ko, and P. L\'evai, Phys. Rev. Lett. {\bf90}, 202302
(2003).
\bibitem{Fries2003PRL}
R. J. Fries, B. M\"uller, C. Nonaka, and S. A. Bass, Phys. Rev.
Lett. {\bf90}, 202303 (2003).
\bibitem{RCHwa2003PRC}
R. C. Hwa and C. B. Yang, Phys.\ Rev.\ C {\bf67}, 034902
(2003).
\bibitem{FLShao2005PRC}
F. L. Shao, Q. B. Xie and Q. Wang, Phys.\ Rev.\ C {\bf71}, 044903
(2005).
\bibitem{TYao2008PRC}
T. Yao, W. Zhou, and Q. B. Xie,  Phys.\ Rev.\ C {\bf78}, 064911 (2008).
\bibitem{CEShao2009PRC}
C. E. Shao, J. Song, F. L. Shao, and Q. B. Xie,  Phys.\ Rev.\ C
{\bf80}, 014909 (2009).
\bibitem{starESC}
Helen Caines (for the STAR Collaboration), arXiv:nucl-ex/0906.0305v1, and references therein.
\bibitem{BES2012nuclex}
Anar Rustamov, arXiv:nucl-ex/1201.4520v1, and references therein.
\bibitem{20A30A2008PRC}
C. Alt {\it et al.} (NA49 Collaboration), Phys. Rev. C {\bf 77},
024903 (2008).
\bibitem{NA49bes2011nuclex}
Marek Gazdzicki (for the NA49 and NA61/SHINE Collaborations), J. Phys. G {\bf 38} 124024 (2011).
\bibitem{CBM2009nuclex}
Dmytro Kresan (for the CBM and NA49 Collaborations), arXiv:nucl-ex/0908.2875v1.
\bibitem{pikp2011nuclex}
Roberto Preghenella (for the ALICE Collaboration), arXiv:nucl-ex/1111.0763v1.
\bibitem{sBpi2012nuclex}
Berndt M\"{u}ller, J\"{u}rgen Schukraft, and Boles{\l}aw Wys{\l}ouch, arXiv:hep-ex/1202.3233v1.
\bibitem{FLShao2007PRC}
F. L. Shao, T. Yao, and Q. B. Xie,  Phys.\ Rev.\ C {\bf75}, 034904
(2007).
\bibitem{pPRC130}
B. I. Abelev {\it et al.} (STAR Collaboration), Phys. Rev. C {\bf 79},
034909 (2009).
\bibitem{kPRC4080158}
S. V. Afanasiev {\it et al.} (NA49 Collaboration), Phys. Rev. C {\bf 66},
054902 (2002).
\bibitem{pPRC200}
S. S. Adler {\it et al.} (PHENIX Collaboration), Phys. Rev. C {\bf 69},
034909 (2004).
\bibitem{plxoPRCSPS}
C. Alt {\it et al.} (NA49 Collaboration), Phys. Rev. C {\bf 78},
034918 (2008), and references therein.
\bibitem{plxoPRC62200}
M. M. Aggarwal {\it et al.} (STAR Collaboration), Phys. Rev. C {\bf 83},
024901 (2011), and references therein.
\bibitem{LPRL130}
K. Adcox {\it et al.} (PHENIX Collaboration), Phys. Rev. Lett. {\bf 89},
092302 (2002).
\bibitem{PLPRC2006}
C. Alt {\it et al.} (NA49 Collaboration), Phys. Rev. C {\bf 73},
044910 (2006).
\bibitem{hypRL200}
J. Adams {\it et al.} (STAR Collaboration), Phys. Rev. Lett. {\bf 98},
062301 (2007).
\bibitem{LPRL1588040}
T. Anticic {\it et al.} (NA49 Collaboration), Phys. Rev. Lett. {\bf 93},
022302 (2004).
\bibitem{MULSPRL130}
J. Adams {\it et al.} (STAR Collaboration), Phys. Rev. Lett. {\bf 92},
182301 (2004).
\bibitem{OSPRL15840}
C. Alt {\it et al.} (NA49 Collaboration), Phys. Rev. Lett. {\bf 94},
192301 (2005).
\bibitem{PPLB62}
I. C. Arsene {\it et al.}, Phys. Lett. B {\bf 677},
267 (2009).
\bibitem{lxoPLB130}
J. Adams {\it et al.} (STAR Collaboration), Phys. Lett. B {\bf 567},
167 (2003).
\bibitem{QBXie1988PRD}
Q. B. Xie and X. M. Liu , Phys.\ Rev.\ D {\bf 38}, 2169  (1988).
\bibitem{RODJPG1995}
Q. Wang and Q. B. Xie, J. Phys. G {\bf 21}, 897 (1995).
\bibitem{PDG2010}
K. Nakamura {\it et al.} (Particle Data Group), J. Phys. G {\bf 37},
075021 (2010).
\bibitem{ZWLinJPG}
Z. W. Lin, J. Phys. G {\bf 38}, 075002 (2011).

\end{thebibliography}
\end{document}